%% file: pricing_matchings.tex
\documentclass[11pt]{article}
\let\chapter\section
\usepackage[numbers]{natbib}
\usepackage[ruled]{algorithm2e}

\usepackage{dsfont}             %
\usepackage{amsmath}
\usepackage{amsthm}
\usepackage{amssymb}
\usepackage{graphicx}
\usepackage{float}
\usepackage{mathabx}

\SetArgSty{textrm}  
\SetAlFnt{\small}
\SetAlCapFnt{\small}
\SetAlCapNameFnt{\small}
\SetAlCapHSkip{0pt}
\IncMargin{-\parindent}

\usepackage{fullpage}
\usepackage{graphicx,amsfonts,amsmath,amssymb,epsfig,color,multicol,pstricks,tikz}
\usepackage{cancel}
\usepackage{paralist}
\usepackage{comment}
\usepackage{url}
\usepackage{hyperref}
\usetikzlibrary{decorations.markings}
\usetikzlibrary{patterns,snakes}
\pgfdeclarelayer{background}
\pgfsetlayers{background,main}
\input{pream}

\newcommand{\itemD}{\text{item-dependent}}

\newcommand{\MyFrame}[1]{\noindent \framebox[\textwidth]{ \begin{minipage}{0.97\textwidth} #1 \end{minipage}}}%

\newcommand{\fp}[0]{\textsf{Price-Items}}
\newcommand{\sapb}[0]{\textsf{Price\text{-}Super\text{-}Additive}}
\newcommand{\pricekdemand}[0]{\textsf{Price\text{-}\textbf{k}{-}Demand}}
\newcommand{\rep}{\Re_{\geq 0}}

\newcommand{\OPT}[0]{\mathsf{OPT}}
\newcommand{\SW}[0]{\textsf{SW}}

\newcommand{\vect}[1]{\ensuremath{\mathbf{#1}}}

\newcommand{\agents}{N}

\newcommand{\goods}{I}
\newcommand{\nitem}{m}

\newcommand{\good}{{\textsf{good}}}

\newcommand{\local}[0]{\textbf{Local}}

\newcommand{\val}{v}
\newcommand{\vals}{\vect{\val}}
\newcommand{\ks}{\vect{k}}

\newcommand{\vali}[1][i]{{\val_{#1}}}

\newcommand{\util}{u}

\newcommand{\utili}[1][i]{\util_{#1}}

\newcommand{\price}{p}
\newcommand{\prices}{\vect{\price}}
\newcommand{\pricei}[1][i]{{\price_{#1}}}

\newcommand{\alloc}{x}
\newcommand{\allocs}{\vect{\alloc}}

\newcommand{\alloci}[1][i]{{\alloc_{#1}}}

\newcommand{\demand}{D}

\newcommand{\demandi}[1][i]{{\demand_{#1}}}

\newcommand{\bundles}{\mathcal{B}}
\newcommand{\bundle}{B}

\newcommand{\Alg}{\text{Alg}}
\newcommand{\itemval}{w}

\newcommand{\out}{\mathrm{out}}

\newcommand{\mpr}{M_{>t}}
\newcommand{\vmod}{v_i^{(b)}}
\newcommand{\valsmod}{\vals^{(b)}}

\newtheorem{lemma}{Lemma}

\newcommand*\samethanks[1][\value{footnote}]{\footnotemark[#1]}

\begin{document}


\title{The Invisible Hand of Dynamic Market Pricing}

\author{Vincent Cohen-Addad \thanks{Sorbonne Universit\'es, UPMC Univ Paris 06, CNRS, LIP6, Paris, France} \\ \texttt{vcohenad@gmail.com}  
	\and Alon Eden \thanks{Tel-Aviv University, Israel} \samethanks[4] \\ \texttt{alonarden@gmail.com}  
	\and Michal Feldman \thanks{Tel Aviv University and Microsoft Research, Israel} \thanks{The work of M. Feldman and A. Eden was partially supported by the European Research Council under the European Union's Seventh Framework Programme (FP7/2007-2013) / ERC grant agreement number 337122.} \\ \texttt{michal.feldman@cs.tau.ac.il}
	\and Amos Fiat \samethanks[2] \thanks{This work was done in part while the A. Fiat was visiting the Simons
		Institute for the Theory of Computing.}\\ \texttt{fiat@tau.ac.il}}





\maketitle

\begin{abstract}
  Walrasian prices, if they exist, have the property that one can assign every buyer some bundle in her demand set, such that the resulting assignment will maximize social welfare. Unfortunately, this assumes carefully breaking ties amongst different bundles in the buyer demand set. Presumably, the shopkeeper cleverly convinces the buyer to break ties in a manner consistent with maximizing social welfare. Lacking such a shopkeeper, if buyers arrive sequentially and simply choose some arbitrary bundle in their demand set, the social welfare may be arbitrarily bad. In the context of matching markets, we show how to compute dynamic prices, based upon the current inventory, that guarantee that social welfare is maximized. Such prices are set without knowing the identity of the next buyer to arrive. We also show that this is impossible in general (e.g., for coverage valuations), but consider other scenarios where this can be done. We further extend our results to Bayesian and bounded rationality models.
\end{abstract}

\section{Introduction}

A remarkable property of Walrasian pricing is that it is possible to match buyers to bundles, such that every buyer gets a bundle in her demand set (i.e., a set of items $S$ maximizing $v_i(S) - \sum_{j\in S} p_j$), and the resulting allocation maximizes the social welfare, $\sum_i v_i(S_i)$ ($S_i$ being the bundle allocated to buyer $i$). However, Walrasian prices cannot coordinate the market alone; it is critical that ties be broken appropriately, in a coordinated fashion.

Consider the following scenario: two items, $a$ and $b$, and two unit demand buyers, Alice and Bob. Alice has value $R \gg 1$ for item $a$ and value one for item $b$,  Bob has value one for each of the two items $a$ and $b$. There are many Walrasian pricings in this setting, for example a price of $R-1$ for item $a$ and 0 for item $b$.  Indeed, assigning item $a$ to Alice and item $b$ to Bob under these prices maximizes simultaneously the individual utility of each buyer and the social welfare.

However, in real markets, buyers often arrive sequentially, in some unknown order, and get no guidance as to how to break ties.
For these prices, ($p(a)=R-1$ and $p(b)=0$), if Bob arrives first then he will indeed choose item $b$, leaving item $a$ for Alice to purchase, resulting in a social welfare maximizing allocation.
If, however,
Alice arrives first, she has equal utility ($=1$)
for both $a$ and $b$ and may select item $b$, so Bob will walk away without purchasing any item, which results in social welfare $1$, compared with the optimal social welfare of $R+1$.
We furthermore remark that setting prices of $p(a)=R$ and $p(b)=1$, which are also Walrasian prices, could result in both Alice and Bob walking away, and resulting in zero social welfare.

One may suspect that we choose the wrong Walrasian pricing. It is known that in matching markets the minimal Walrasian prices coincide with VCG payments \cite{Leonard83}, and they are also the outcome of natural ascending auctions for matching markets \cite{CK81}. In this example the minimal Walrasian prices are to charge zero for both item $a$ and item $b$. Indeed, if Alice arrives first, she will choose item $a$, and when Bob arrives he will choose item $b$, and this is the social welfare maximizing allocation. However, if Bob arrives first, he will be indifferent between the two items and may choose item $a$ --- again --- this achieves a social welfare of $2$ compared with the optimal social welfare of $R+1$.
In fact, one can show that minimal Walrasian prices always induce ties in demand \cite{HsuMRRV15}.
Moreover, there exist markets that admit unique Walrasian prices, yet may achieve zero welfare.
For example, consider a single item valued at 1 by both Alice and Bob. The unique Walrasian price is 1, which may result in both buyers walking away without purchasing the item.

In fact, we can show that no static prices (and thus no Walrasian prices) can give more than 2/3 of the social welfare for buyers that arrive sequentially. Consider unit demand buyers Alice, Bob, and Carl, and items $a$, $b$, and $c$. Alice values $a$ and $b$ at one, and has zero value for $c$; symmetrically, Bob values $b$ and $c$ at one and $a$ at zero, and Carl values $c$ and $a$ at one, and $b$ at zero. We term such unit-demand valuations, where all individual values are either 0 or 1, binary unit-demand valuations. A two line proof shows that no static pricing scheme, $p(a)$, $p(b)$, and $p(c)$ can achieve more than 2/3 of the optimal social welfare for this valuation profile. Assume all prices are strictly less than one, and assume, without loss of generality, that $p(a)\geq p(b) \geq p(c)$. Now, Alice arrives and chooses item $b$, Carl arrives and chooses item $c$, and finally Bob arrives --- but there are no items left for which Bob has  a non zero valuation. Note that if $p(a)\geq 1$ then item $a$ will not be sold as whomever is to buy it may decide simply to walk away, the same holds for items $b$ and $c$ so assuming that all prices are strictly less than one holds without loss of generality, given that one assumes that the prices achieve $\geq 2/3$ of the optimal social welfare.

However, consider the following twist, which changes the prices after the first buyer arrives. In the scenario above, when Alice arrives first and chooses (without loss of generality) item $a$, change the prices so that Bob will choose $b$ and Carl will choose $c$. This is easily done by setting new prices $p'(b)<p'(c)$. Irrespective of whomever arrives after Alice, the prices will ensure that all items get sold and social welfare be maximized.

Obtaining optimal social welfare is trivial via dynamic pricing if the pricing mechanism knew which buyer was to arrive next. The dynamic pricing mechanism could make use of infinite prices to reduce the choices available to incoming buyer so that  only a bundle consistent with optimal social welfare can be selected. The key difficulty arises because the prices need to be set {\sl before} the preferences of the next buyer to arrive are known.

Thus, this paper studies the issues of static and dynamic pricing for sequentially arriving buyers. Our main result is the following:

\vspace{0.1in}

\noindent {\bf Main Theorem:}
For any matching market ({\sl i.e.}, unit demand valuations), we give a poly-time dynamic pricing scheme that achieves the optimal social welfare, for any arrival order and irrespective of any tie breaking chosen by the buyers. We give a combinatorial algorithm to compute such prices\footnote{We are grateful to an anonymous reviewer who pointed out that this result can also be obtained via an LP formulation of the problem and using LP-duality.}.

\vspace{0.1in}

We show that the existence of Walrasian prices does not, by itself, imply that there exist dynamic pricing schemes that optimize social welfare. In particular, we give an example (Section \ref{sec:coverage}) of a market with coverage valuations (a strict subclass of submodular valuations), which has a unique optimal solution, and where Walrasian prices do exist, and yet no pricing scheme (static or dynamic) can get the optimal social welfare.

\vspace{0.1in}

We offer some remedies for this impossibility result.

\begin{itemize}
\item We show that a market with gross substitutes valuations that has a unique optimal allocation always admits a {\sl static item pricing} scheme that achieves the optimal welfare (Section \ref{sec:staticunique}).
\item Moreover, while full efficiency is in general impossible, we argue that for {\em any} profile of valuations, there exists a static pricing scheme that achieves at least a half of the optimal social welfare. This result can be viewed as a generalization of the Combinatorial Walrasian Equilibrium of \cite{feldman2013combinatorial}. In fact we adapt the static bundle prices computed in \cite{feldman2015combinatorial} for Bayesian agents to achieve the one half guarantee of the optimal social welfare, for any class of valuations.
\item We identify additional classes of valuations that admit dynamic pricing schemes that obtain the optimal social welfare: (1) where buyer $i$ seeks up to $k_i$ items, and valuations depend on the item, and (2) for superadditive valuations.
\end{itemize}

The following remark is in order. Gross substitutes valuations are known to be the frontier for the guaranteed existence of a Walrasian equilibrium \cite{gul1999walrasian}. They are also the frontier with respect to computational tractability \cite{Nisan06thecommunication}: one can compute the allocation that maximizes social welfare in polynomial time. The following open problem emerges from our work: are gross substitutes valuations also the frontier for achieving optimal welfare via a dynamic pricing scheme? In other words: Does every market with gross substitutes valuations admit a dynamic pricing scheme that achieves optimal social welfare?

An additional problem that remains open concerns the power of static pricing. As stated above, no static prices can give more than 2/3 of the optimal social welfare, even in the case of unit-demand buyers. On the other hand, it is known that there exist static prices that achieve at least a half of the optimal social welfare, even in the case of submodular (or XOS) valuations, and even in a Bayesian setting \cite{feldman2015combinatorial}.
While 1/2 has been shown to be tight for Bayesian settings, the negative example does not carry over to settings with full information. A natural open problem is therefore closing the gap between 1/2 and 2/3 of the fraction of the optimal welfare that can be achieved by static pricing in full information settings. This problem is open even for the class of binary unit-demand valuations. Recently, it has been shown that static prices can achieve more than a 0.51 fraction of the optimal welfare in the case of binary unit-demand valuations \cite{Maxmin}, leaving a gap between ~0.51 and 2/3 for binary unit-demand valuations, and a gap between 1/2 and 2/3 for general unit-demand valuations (and more general ones).


\subsection{Related Work}

 Walrasian equilibrium, where prices are such that optimal social welfare is achieved, and the market clears, given appropriate tie-breaking of preferences in the demand set dates back to 1874 \cite{walras1874}. The existence of Walrasian prices for matching markets and more generally for gross substitutes valuations appears in \cite{kelso1982job,gul1999walrasian}. The efficiency guarantee for the minimal Walrasian prices was considered in \cite{HsuMRRV15} for both matching markets and matroid based valuations.\footnote{The class of matroid based valuations are conjectured to be equivalent to the class of gross substitutes valuations by \cite{OstrovskyTE1840}.} Interestingly, the question of finding a Walrasian equilibrium without ties in the demand for the case of gross substitutes valuations with a unique optimum was also asked by \cite{LemeW15} in the quite different context of efficiently computing a Walrasian equilibrium given access to an aggregate demand oracle.
 For a survey on gross substitutes valuations, see \cite{leme2014gross}.


 The use of pricing schemes in order to maximize welfare in combinatorial auctions was considered by \cite{feldman2015combinatorial} where they show static item prices can obtain $1/2$ of the optimal social welfare for buyers arriving via a Bayesian process, with XOS valuations. This result was later extended and generalized to a general framework of stochastic online welfare maximization by \cite{DuettingFKL17}. Improved bounds were shown for both the Bayesian and full information case by \cite{EzraFRS17} when the items are identical. The question of whether static item prices can get more than half of the optimal welfare was tackled by \cite{Maxmin}, where they showed that for binary unit-demand valuation, the answer is affirmative. The use of bundle prices was considered by \cite{feldman2013combinatorial} in order to circumvent the impossibilities of achieving Walrasian equilibrium when item prices are in use.

 The performance of posted price mechanisms was also studied under the objective of maximizing revenue in Bayesian settings.
 When a single item is for sale, \cite{HajiaghayiKS07} noticed that results in the prophet inequalities settings translate to nearly optimal mechanism.
 \cite{GuruswamiHKKKM05} considered the question of finding envy-free prices that simultaneously maximize revenue in the full information setting, and show a logarithmic approximation for this objective for unit-demand and single-minded bidders. Since a better approximation than to a polylog factor is impossible, \cite{Chawla07} considered a Bayesian setting where a single unit demand buyer is sampled from a product distribution over the items, and showed how to price the items in order to get a constant factor approximation. The result was later extended by \cite{Chawla10} to multiple unit demand buyers using dynamically adjusted prices.

Dynamic posted prices were considered in a broader setting than auctions by \cite{cohen2015pricing}, where they show posted price mechanisms can achieve a nearly optimal competitive ratio for a host of cost minimization problems. Lately, there has been a growing interest in analyzing the performance of posted price mechanisms for various scheduling objectives such as makespan minimization \cite{FeldmanFR17}, maximum flow time minimization \cite{ImMPS17} and sum of weighted completion times minimization \cite{EdenFFT18}.

Our work is also related to the rich literature on online bipartite matching. The first work to consider a similar setting was the seminal paper by \cite{karp1990optimal} where side $A$ of a bipartite graph is known in advance, while the vertices in side $B$ arrive sequentially, each vertex with its adjacent edges, and the algorithm's goal is to match each incoming vertex while maximizing the size of the matching. \cite{karp1990optimal} gave an optimal randomized $1-1/e$ approximation to the size of the matching. An analogy of their algorithm to posted prices was given by \cite{EdenFFS18}. For a broad overview of online matching problems and results in this area, we refer to the survey in \cite{Mehta13}.


\subsection{Paper Structure}

In Section \ref{sec:model} we describe our model, and define different types of markets and pricing schemes.

In Section \ref{sec:unitdemand} we present our main result, which is a dynamic pricing scheme that achieves the optimal social welfare, irrespective of the order of agent arrival and the form in which agents break ties.

In Section \ref{sec:unique} we study markets that admit a unique optimal allocation.
In Section \ref{sec:staticunique} we show that our main result can be extended to a market with gross substitutes valuations, if the optimal allocation is unique. In particular, one can always compute static item prices that lead to optimal welfare.
These prices are Walrasian prices (but not any Walrasian prices).
In Section \ref{sec:coverage} we show that extending this result beyond gross substitutes valuations is not always possible, even with dynamic prices. In particular, we show an example of a market with coverage valuations (a subset of submodular valuations) that admits a unique optimal allocation, as well as a Walrasian equilibrium, but where no dynamic prices can lead to optimal welfare.

In Section \ref{sec:bundle-prices} we study the power of bundle prices.
In Section \ref{sec:general-val} we show that if one allows to assign prices to bundles (as opposed to items), then a half of the optimal social welfare can always be achieved (i.e. for arbitrary monotone valuations), even with static prices.
The construction of prices uses ideas from \cite{feldman2013combinatorial,feldman2015combinatorial}.
In Sections \ref{sec:superadditive} and \ref{sec:k_i_demand_vertex-weighted} we present two classes of valuations for which static bundle prices can lead to optimal social welfare. These are super-additive valuations and some variant of unit-demand valuations.


%

\section{Model and Preliminaries}

\label{sec:model}

\input{model}

\section{Optimal Dynamic Pricing Scheme for Matching Markets} \label{sec:unitdemand}

\input{unit_demand}

\section{Unique Optimum}\label{sec:unique}
In this section, we inspect the case where the social maximizing allocation is unique. We first show that in this case, an optimal dynamic bundle-pricing scheme implies an optimal static bundle-pricing scheme:

\begin{obs}
	Let $\vals=(v_1,\ldots,v_n)$, where $v_i:2^\goods \mapsto \rep$, and let $\langle \vals, \goods\rangle$ be an instance where $\mathcal{B}=\{B_1,\ldots,B_n\}$ is the unique partition of items that maximizes social welfare. If there exists an optimal dynamic bundle-pricing scheme, then there must exist an optimal static bundle-pricing scheme.
\end{obs}

\begin{Proof}
	Let $\price_1:\mathcal{B}\rightarrow \rep$ be the initial prices the optimal dynamic pricing scheme gives to the bundles. We claim that sticking to these prices throughout the process guarantees an optimal allocation as well. Without loss of generality, assume that agents with lower index arrive earlier and that the $i$-th agent to arrive is the first agent whose choice $X\neq \{B_i\}$ (could be that $X=\{B_j\}$, $j\neq i$, could be that $x=\{B_i,B_j, \ldots\}$, $j\neq i$, and could be that $X=\emptyset$).
	
	It must be the case that $u_i(\price_1,X)\geq u_i(\price_1,B_i)$. Therefore, if this agent arrives first, she is not guaranteed to take $\{B_i\}$ since this not the unique bundle that maximizes her utility. This contradicts the optimality of the dynamic pricing scheme.
\end{Proof}

\subsection{Optimal Welfare for Gross Substitutes Valuations}\label{sec:staticunique}
\input{static_unique}

\subsection{Impossibility Result for Coverage Valuations}\label{sec:coverage}
\input{coverage}

\section{The Power of Bundle Prices} \label{sec:bundle-prices}

Recall that a bundle pricing algorithm partitions the items into bundles, $\bundles=\{\bundle_1,\ldots,\bundle_k\}$ ($\bigcup_i \bundle_i=\goods$), and assigns a price to every bundle in $\bundles$. In this section we show how pricing bundles can help us in getting optimal and approximately optimal welfare guarantees.

\subsection{Bundle Prices for General Valuations}
\label{sec:general-val}
\input{static_half}
\subsection{Bundle Prices for Super-additive Valuations - New}\label{sec:superadditive}
\input{superadd_new.tex}

\subsection{Bundle Prices for $\ks$-Demand Item-Dependent Valuations}\label{sec:k_i_demand_vertex-weighted}
\input{k_i_demand_vertex-weighted}


\subsection*{Acknowledgements}
We are grateful to Claire Mathieu and Orit Raz for helpful discussions.
This work was partially supported by the European Research Council under the European Union's Seventh Framework Programme (FP7/2007-2013) / ERC grant agreement number 337122, and by the Israel Science Foundation (grant number 317/17).


\bibliographystyle{abbrv} 
\bibliography{pricing}

\appendix

\section{No Static Prices for the Running Example (Figure~\ref{fig:all-phases})}

\begin{lem}
  \label{lem:nostaticprices}
  There is no static pricing scheme for the running example that achieves optimal welfare.
\end{lem}

\begin{Proof}
  Note that in any welfare maximizing allocation for the example, all items should be allocated.
  We consider pricing
  Let $D_{\text{Alice}}$, $D_{\text{Bob}}$ and, $D_{\text{Carl}}$ denote the demand sets of Alice, Bob, and Carl, respectively, under pricing $p$.

Suppose that $|D_{\text{Alice}}| = 2$. Then if $c \notin D_{\text{Bob}} \cup D_{\text{Carl}}$, then $D_{\text{Bob}} = \{b\}$ and $D_{\text{Carl}} = \{a\}$.
We consider the following sequence: Carl arrives first and takes a, Bob arrives second and takes b and so $c$ is not picked, a contradiction.
So suppose $c$ belongs to $D_{\text{Bob}}$, then we consider the following order of arrival:
Bob arrives first and takes $c$, Alice arrives second and takes $a$ and so $b$ is not picked, a contradiction.
Similarly if $c \in D_{\text{Carl}}$, we consider the arrival where Carl arrives first and takes $c$, Alice arrives second and takes $b$ and so
at least one of $a$ or $d$ is not picked, a contradiction.
Symmetrically, the above argument applies to the cases where $|D_{\text{Bob}}| = 2$ or $|D_{\text{Carl}}| = 2$.

Then, suppose that $|D_{\text{Alice}}|,|D_{\text{Bob}}|,|D_{\text{Carl}}| = 1$.
Suppose first that $D_{\text{Alice}}=\{a\}$ and so, $D_{\text{Bob}} = \{b\}$ and $D_{\text{Carl}}=\{c\}$.
Then $6-p(a) > 12-p(b)$, $8-p(b) > 8-p(c)$ and $10 - p(c) > 4-p(a)$.
Combining we obtain, $6 + p(a) > p(c) > p(b) > 6 + p(a)$ a contradiction.
Suppose then that $D_{\text{Alice}}=\{b\}$ and so, $D_{\text{Bob}} = \{c\}$ and $D_{\text{Carl}}=\{d\}$.
Then $12-p(b) > 6-p(a)$, $8-p(c) > 8-p(b)$ and $4 - p(a) > 10-p(c)$.
Combining, $6 + p(a) > p(b) > p(c) > 6 + p(a)$, a contradiction.
The assertion of the lemma follows.
\end{Proof}











\end{document}

%% file: pream.tex
\renewcommand{\paragraph}[1]{{\protect\vspace{8pt}\noindent\sc{#1}}}
\usepackage{multicol}
\usepackage[nottoc]{tocbibind}

\usepackage[noend]{algpseudocode}

\usepackage{dsfont}
\usepackage{enumitem}

\newlength{\saveparindent}
\newlength{\saveparskip}
\newcommand{\BE}{\begin{enumerate}} \newcommand{\EE}{\end{enumerate}}
\newcommand{\BI}{\begin{itemize}} \newcommand{\EI}{\end{itemize}}
\newcommand{\BDes}{\begin{description}}\newcommand{\EDes}{\end{description}}
\newtheorem{alg}{Algorithm}
\newcommand{\BA}{\begin{alg}} \newcommand{\EA}{\end{alg}}
 \newtheorem{thm}{Theorem}[section]            
\newcommand{\BT}{\begin{thm}} \newcommand{\ET}{\end{thm}}

\newtheorem{lem}[thm]{Lemma} 
\newcommand{\BL}{\begin{lem}} \newcommand{\EL}{\end{lem}}

\newtheorem{clm}[thm]{Claim}
\newcommand{\BCM}{\begin{clm}} \newcommand{\ECM}{\end{clm}}

\newtheorem{techcor}[thm]{Corollary}
\newcommand{\BCo}{\begin{techcor}} \newcommand{\ECo}{\end{techcor}}

\newtheorem{obs}[thm]{Observation}

\newtheorem{Conc}[thm]{Conclusion}
\newcommand{\BCONC}{\begin{Conc}} \newcommand{\ECONC}{\end{Conc}}

\newtheorem{Obs}[thm]{Observation}
\newcommand{\BOBS}{\begin{Obs}} \newcommand{\EOBS}{\end{Obs}}

\newtheorem{Exmp}[thm]{Example}
\newcommand{\BEXM}{\begin{Exmp}} \newcommand{\EXMP}{\end{Exmp}}

\newtheorem{cor} [thm] {Corollary}      
\newcommand{\BC}{\begin{cor}} \newcommand{\EC}{\end{cor}}
\newtheorem{prop}[thm]{Proposition}     
\newcommand{\BP}{\begin{prop}} \newcommand {\EP}{\end{prop}}
\newtheorem{conj} {Conjecture}      
\newcommand{\BCJ}{\begin{conj}} \newcommand{\ECJ}{\end{conj}}

\newtheorem{fact}[thm]{Fact}
\newcommand{\BF}{\begin{fact}} \newcommand{\EF}{\end{fact}}
\newtheorem{assumption}[thm]{Assumption}
\newcommand{\BAs}{\begin{assumption}} \newcommand{\EAs}{\end{assumption}}

\newtheorem{defn}{Definition}[section]         
\newcommand{\BD}{\begin{defn}} \newcommand{\ED}{\end{defn}}

\def\FullBox{\hbox{\vrule width 8pt height 8pt depth 0pt}}
\newcommand{\QED}{\;\;\;\FullBox}
\newenvironment{Proof}{\noindent{\bf Proof:~~}}{\hfill\QED}
\newcommand{\BPF}{\begin{Proof}} \newcommand {\EPF}{\end{Proof}}
\newenvironment{proofof}[1]{\noindent{\bf Proof of {#1}:~~}}{\(\hfill\QED\)}
\newcommand{\BPFOF}{\begin{proofof}} \newcommand {\EPFOF}{\end{proofof}}

\newenvironment{smallproof}{\noindent{\bf Proof sketch:~~}}{\(\QED\)}
\newcommand{\bpf}{\begin{smallproof}} \newcommand{\epf}{\end{smallproof}}

\newcommand{\BEQ}{\begin{equation}} \newcommand{\EEQ}{\end{equation}}
\newcommand{\BEQN}{\begin{eqnarray}}\newcommand{\EEQN}{\end{eqnarray}}

\newcommand{\eps}{\epsilon}

\newcommand*{\rom}[1]{\expandafter\@slowromancap\romannumeral #1@}

\newcommand{\D}{{\rm D}}

\newcommand{\dist}{{\rm dist}}

\newcommand{\argmax}{{\rm argmax}}

\newcommand\abs[1]{\left|#1\right|}

%% file: model.tex
Our setting consists of a set $\goods$ of $\nitem$ indivisible items
and a set of $n$ buyers that arrive sequentially in some arbitrary order.

Each buyer has a valuation function $\vali : 2^\goods \to \Re_{\geq 0}$ that indicates her value for
every set of objects, and a buyer valuation profile is denoted by $\vals=(\val_1,\dotsc,\val_n)$. We assume valuations are monotone non-decreasing and normalized (i.e., $\vali(\emptyset) = 0$). We use $\vali(A|B)=\vali(A\cup B)-\vali(B)$ to denote the marginal value of bundle $A$ \textit{given} bundle $B$.
An {\em allocation} is a vector of disjoint sets $\allocs = (\alloc_1, \dotsc, \alloc_n)$, where $\alloci$ denotes the bundle associated with buyer $i \in [n]$ (note that it is not required that all items are allocated).
The {\em social welfare} (\SW) of an allocation $\allocs$ is $\SW(\allocs,\vals)=\sum_{i=1}^{n}\vali(\alloci)$, and the optimal welfare is denoted by OPT($\vals$).
When clear from context we omit $\vals$ and write $\SW(\allocs)$ and OPT for the social welfare and optimal welfare, respectively.

An {\em item pricing} is a function $\prices:\goods \rightarrow \Re^{\geq 0}$ that assigns a price to every item. The price of item $j$ is denoted by $\price(j)$.
Given an item pricing, the \emph{utility} that buyer $i$ derives from a set of items $S$ is $\utili(S, \prices) = \vali(S) - \sum_{j \in S}\price(j).$ The {\em demand correspondence} $\demandi(\goods, \prices)$ of buyer $i$ contains the sets of objects that maximize buyer $i$'s utility; i.e.,
$\demandi(\goods, \prices) = \argmax_{S \subseteq \goods} \utili(S, \prices)$.

A {\em bundle pricing} is a tuple $(\bundles,\prices)$, where $\bundles=\{\bundle_1,\ldots,\bundle_k\}$ is a partition of the items into bundles (where $\bigcup_i \bundle_i=\goods$ and for every $i\neq j$, $\bundle_i \cap \bundle_j=\emptyset$), and $\price:\bundles \rightarrow \Re^{\geq 0}$ is a function that assigns a price to every bundle in $\bundles$. The price of bundle $\bundle_j$ is denoted $\price(\bundle_j)$.
Given a bundle pricing $(\bundles,\prices)$, the utility that buyer $i$ derives from a set of bundles $S$ is $\utili(S, \prices) = \vali(S) - \sum_{\bundle_j \in S}\price(\bundle_j).$ 
The {\em demand correspondence} $\demandi(\bundles, \prices)$ of buyer $i$ contains the sets of bundles that maximize buyer $i$'s utility; i.e., $\demandi(\bundles, \prices) = \argmax_{S \subseteq \bundles} \utili(S, \prices)$.

We consider several types of pricing schemes: {\em static item pricing}, {\em dynamic item pricing}, {\em static bundle pricing}, and {\em dynamic bundle pricing}.

In static pricing schemes, prices are assigned (to items or bundles) initially, and never change then. In contrast, in dynamic pricing schemes, new (item or bundle) pricing may be set before the next buyer arrives. Item pricing schemes assign prices to items, whereas bundle pricing schemes partition the items to bundles and assign prices to bundles that are elements of the partition. Thus, the four types of pricing schemes are described as follows.

\vspace{0.1in}
\noindent {\bf Static Item Pricing Scheme:}
\begin{enumerate}
\setlength{\itemsep}{5pt}
\setlength{\parskip}{0pt}
\setlength{\parsep}{0pt}
\item Item prices, $\prices$, are determined once and for all.
\item Buyers arrive in some arbitrary order, the next buyer to arrive chooses a bundle in her demand set among the items not already allocated (and pays the sum of the corresponding prices).
\end{enumerate}

\noindent {\bf Static Bundling Pricing Scheme:}
\begin{enumerate}
\setlength{\itemsep}{5pt}
\setlength{\parskip}{0pt}
\setlength{\parsep}{0pt}
\item Bundles, and their prices, $(\bundles,\prices)$, are determined once and for all.
\item Buyers arrive in some arbitrary order, the next buyer to arrive chooses a set of bundles in her demand set among the bundles not already allocated (and pays the sum of the corresponding prices).
\end{enumerate}

\noindent {\bf Dynamic Item Pricing Scheme:}
\begin{itemize}
\setlength{\itemsep}{5pt}
\setlength{\parskip}{0pt}
\setlength{\parsep}{0pt}
\item Before buyer $t=1,\ldots,n$ arrives (and after buyer $t-1$ departs, for $t>1$):
\begin{enumerate}
\item Item prices, $\prices_t$, are set (or reset) before buyer $t$ arrives, prices are set for those items that have not been purchased yet.
\item When buyer $t$ arrives she purchases a set of items $S$ in her demand among the items not already allocated (and pays the sum of the corresponding prices according to $\prices_t$).
\end{enumerate}
\end{itemize}

\noindent {\bf Dynamic Bundle Pricing Scheme:}
\begin{itemize}
\setlength{\itemsep}{5pt}
\setlength{\parskip}{0pt}
\setlength{\parsep}{0pt}
\item Before buyer $t=1,\ldots,n$ arrives (and after buyer $t-1$ departs, for $t>1$):
\begin{enumerate}
\item A partition into bundles and bundle prices, $(\bundles,\prices_t)$, is determined for the items that have not been purchased yet.
\item When buyer $t$ arrives she purchases a set of bundles $S$ in her demand set among the bundles on sale (and pays the sum of the corresponding prices according to $\prices_t$).
\end{enumerate}
\end{itemize}

We say that a pricing scheme achieves optimal (respectively, $\alpha$-approximate) social welfare if for any arrival order and any manner in which agents may break ties, the obtained social welfare is optimal (resp., at least an $\alpha$ fraction of the optimal welfare).

\subsection{Classes of Valuations}\label{sec:valuations}
This paper introduces a host of dynamic and static pricing scheme for a variety of well studied valuation classes. In the following, we formally define these classes.

\vspace{0.1in}
\noindent {\bf Unit demand valuations:} A unit demand valuation $v:2^\goods\rightarrow \rep$ associates each item $j\in \goods$ with a value $v^j$, and values a bundle $S\subseteq \goods$ by the maximum valued item in the bundle, that is, $v(S)=\max_{j\in S}v^j$.  

\vspace{0.1in}
\noindent {\bf Gross substitutes valuations\cite{kelso1982job}:} A valuation function $v_i$ is gross substitutes if for every price vector $\prices$, a set $S\in \demandi(\goods,\prices)$ and price vector $\prices'\geq \prices$, there exists some set of items $T\subseteq \goods$ such that $$\{j\in S:p(j)=p'(j)\}\cup T\in \demandi(\goods,\prices').$$  

\vspace{0.1in}
\noindent {\bf Coverage valuations:} A coverage valuation $v$ is defined by a ground set $E$ of elements, a weight function on the elements $w:E\rightarrow \rep$ and a mapping $m:\goods\rightarrow 2^E$ from items to sets of elements. For a given set of items $S$, let $E_S=\bigcup\limits_{j\in S}m(j)$ be the set of elements covered by the items in $S$. The value of $S$ is $v(S)=\sum\limits_{e\in E_S}w(e).$

\vspace{0.1in}
It is easy to see that every unit demand valuation is also gross substitutes and coverage. On the other hand, there are gross substitutes valuations that are not coverage and vice versa. We also note that both gross-substitutes and coverage valuations are strict subsets of submodular valuations.

\vspace{0.1in}
\noindent {\bf Super-additive valuations:} A valuation $v$ is said to be super-additive if for every two disjoint sets of items $A,B\subseteq \goods$, $v(A\cup B)\geq v(A)+v(B)$. 

\vspace{0.1in}
In Section \ref{sec:k_i_demand_vertex-weighted} we extend our results to a less studied class of valuations which we term $\ks$-demand $\itemD$. We define the class formally in the corresponding section. We note that this class is contained in the class of gross substitutes. It is not contained in and does not contain the class of unit demand valuations. 

%% file: unit_demand.tex
In this section we consider unit demand valuations (matching markets). Every agent seeks one item, and may have different valuations for the different items. Whereas this setting admits Walrasian prices, such prices are not applicable to the setting where agents arrive sequentially, in an unknown order, and choose an arbitrary item in their demand set. 

We now describe a dynamic item pricing scheme for matching markets that maximizes social welfare --- the sum of buyer valuations for their allocated items is maximized. The process we consider is as follows:

\vspace*{-4pt}
\begin{itemize}
\setlength{\itemsep}{1pt}
\setlength{\parskip}{0pt}
\setlength{\parsep}{0pt}
\item The valuations of the buyers are known.
\item The buyers arrive in some arbitrary order unknown to the pricing scheme.
\item Prices are posted, they may change after a buyer departs but cannot depend upon the next buyer.
\end{itemize}

\begin{figure}
    \includegraphics[width=300bp]{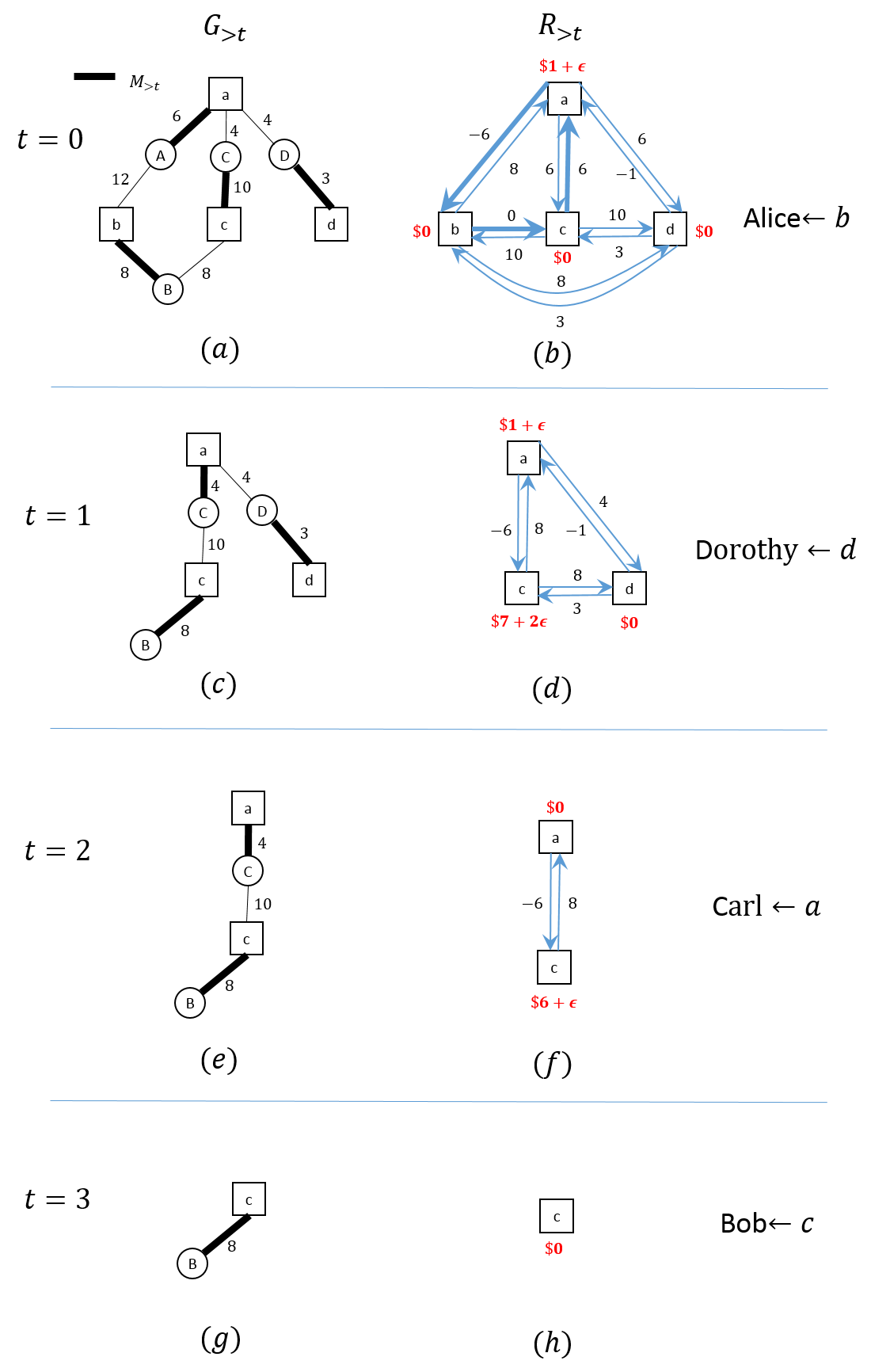}
    \caption{An Illustration of our Running Example: Phases $t=0,1,2,3$ of our running example. Squares represent items and circles represent buyers. Every row represents a phase in the process, where a single buyer arrives. On the left one sees the graph representing the valuations of the remaining buyers and items (graphs labeled $(a)$, $(b)$, $(c)$ and $(d)$, where thick edges represent a maximum weight matching in the graph. Graphs labeled $(b)$, $(d)$, $(f)$ and $(h)$ give the graphs $R_{>t}$ from which the dynamic are computed. Directed cycles of length 0 (if any) are represented by thick edges, after they are discarded, prices are computed  via Algorithm $\fp$. On the very right one sees the next buyer to arrive as well as the item she chooses (based upon the pricing, and breaking ties for the sake of this example.}
    \label{fig:all-phases}
\end{figure}

\setlength{\fboxrule}{2pt}
  \noindent \fbox{\noindent\makebox[0.95\textwidth][c]{\begin{minipage}{0.9\textwidth} {\bf [Running example]} To illustrate the process, we consider a running example of a matching market, buyers Alice, Bob, Carl, and Dorothy, items $a$, $b$, $c$ and $d$. The valuations are given in Figure~\ref{fig:all-phases}(a), where squares represent buyers, circles represent items, and $A,B,C,D$ stand for Alice, Bob, Carol and Dorothy. The minimal Walrasian pricing is $p(a)=1,p(b)=7,p(c)=7,p(d)=0$.
  Under the minimal Walrasian pricing, or any static pricing, unless ties are broken in a particular way, sequential arrival of buyers will not produce optimal social welfare (as shown in the full version of the paper).

  An example of the use of dynamic pricing that follows from our dynamic pricing scheme is given in Figure~\ref{fig:all-phases}. Every row represents a phase in the process, where a single buyer arrives. The LHS graph in every row represents the valuations of the remaining buyers and items, thick edges represent a maximum weight matching. The RHS graph represents the graph of edges, upon which prices are calculated by Algorithm $\fp$. Directed cycles of length 0 (if any) are represented by thick edges. The arriving buyers along with the items they pick are specified in the right column.\end{minipage}}} \par\setlength{\fboxrule}{0.1pt}

\vspace{0.1in}

The input consists of the graph
$G=(\agents,\goods,\vals)$. $G$ is a complete bipartite weighted graph, where $\agents$ is the set of agents, $\goods$ is the set of items, and for every agent $a\in \agents$ and item $b\in \goods$, the weight of an edge $\langle a, b \rangle$ is the value that agent $a$ gives item $b$, $v_a(b)$ ($v_a:\goods\rightarrow\rep$ is the valuation function for agent $a$).

Without loss of generality, one may  assume that in $G$ we have that $\abs{\goods}\geq \abs{\agents}$, otherwise, we add dummy vertices to the $\goods$ with zero weight edges to the vertices of the $\agents$ side until $\abs{\goods}=\abs{\agents}$. $OPT$ is the weight of the maximum weight matching in $G$ (alternatively, the optimal social welfare). Let $M\subseteq \agents\times \goods$ be some matching in $G$, we define $\SW(M)=\sum_{(a,b)\in M} \val_a(b)$ to be a function that takes a matching and returns the social welfare (value) of the matching.

We now continue to describe the dynamic pricing scheme. At time $t\in 0,\ldots, \abs{\agents}$ (after the $t$-th agent departs), we define the following:
\begin{itemize}
\setlength{\itemsep}{1pt}
\setlength{\parskip}{0pt}
\setlength{\parsep}{0pt}
	\item $M_t\subseteq \agents\times \goods$ is the partial matching consisting of a subset of the first $t$ agents to arrive, and the item of their choice, amongst the items available for sale upon arrival. The size of $M_t$ may be less than $t$. It can indeed be the case that
not all buyers may be matched since their demand set may be empty when they arrive.
	\item $\agents_t\subseteq \agents$ and $\goods_t\subseteq \goods $ are the first $t$ agents to arrive and the items matched to them in the matching $M_t$.
	\item $\agents_{>t}=\agents\setminus \agents_t$ and $\goods_{>t}=\goods\setminus \goods_t$ are the remaining agents (to arrive at some time $>t$) and the items remaining after the departure of the $t$-th agent. Define $G_{>t}$ to be the graph $G$ where agents $\agents_t$ and items $\goods_t$ have been discarded. {\sl I.e.}, $G_{>t}=\left(\agents_{>t}, \goods_{>t}, \vals\right)$.
	\item We define $\price_{t+1}:\goods_{>t}\rightarrow \rep$ to be the prices set by the dynamic pricing scheme after the departure of agent $t$ (but before the arrival of agent $t+1$).  \end{itemize}
To compute the function $\price_{t+1}$ we first construct a so-called ``relation graph", $R_{>t}$, and then perform various computations upon it. The vertices of the relation graph are all items yet unsold, $\goods_{>t}$, the edges and their weights are as follows:
\begin{enumerate}
	\item Compute $M_{>t}\subseteq \agents_{>t}\times \goods_{>t}$, a maximum weight matching of the graph $G_{>t}$ which matches \textbf{all} vertices of $\goods_{>t}$.\footnote{Note that such a maximum weight matching exists because initially $\abs{\agents}\leq \abs{\goods}$, and since every agent takes at most one item, $\abs{\agents_{>t}}\leq \abs{\goods_{>t}}$ continues to hold. Since all edge weights are non-negative, and $G_{>t}$ is a complete bipartite graph, every maximum weight matching can be extended to produce a  matching with  the same weight which matches all of the vertices in $\goods_{>t}$.} For every item $b\in \goods_{>t}$, let $\val_{>t}(b)$ denote the value of item $b$ to the agent matched to item $b$ in the matching $M_{>t}$.\label{step:find_mathcing}
	\item The edges of $R_{>t}$, denoted by $E_{>t}$, are a clique on the vertices $\goods_{>t}$, and their weights  $W_{>t}:E_{>t}\rightarrow \Re$ are computed as follows:
  Let $M_{>t}$ be a maximum weight matching of remaining items and agents as defined above. For every pair $(a,b)\in M_{>t}$, and for every $b'\in \goods_{>t}\setminus \{b\}$ create an edge $\langle b,b'\rangle$. The weight of the edge $\langle b,b'\rangle$, \begin{eqnarray}
  		W_{>t}(\langle b,b' \rangle) =v_a(b)-v_a(b').\label{eq:weight_def}
  \end{eqnarray}
 \end{enumerate}

\setlength{\fboxrule}{2pt}
\noindent \fbox{\noindent\makebox[0.95\textwidth][c]{\begin{minipage}{0.9\textwidth}
 {\bf [Running example]} The initial graph $G_{>0}$ of our running example is given in Figure \ref{fig:all-phases}(a), where a maximum weight matching $M_{>0}$ is indicated by thick edges. The graph $R_{>0}$ is given in Figure \ref{fig:all-phases}(b). For example, the weight of the edge $\langle a,b\rangle$ is $v_{Alice}(a)-v_{Alice}(b)=-6.$
\end{minipage}}}\setlength{\fboxrule}{0.1pt} \par

\vspace{0.1in}

We give the following structural property of $R_{>t}$:
\begin{lemma}
	There are no directed cycles of negative weight in $R_{>t}$.\label{lem:no_neg_cycles}
\end{lemma}
\begin{Proof}
	Assume there exists a negative cycle of length $\ell$. Assume the cycle is comprised of $\langle b_1,b_2\rangle,\langle b_2,b_3\rangle,\ldots,\langle b_{\ell-1},b_{\ell}\rangle,\langle b_{\ell},b_1\rangle$. This cycle corresponds to a cycle of alternating edges in $G_{>t}$ $\left(b_1,a_1\right)\left(a_1,b_2\right),\left(b_2,a_2\right)\ldots\left(a_{\ell-1},b_{\ell}\right),\left(b_{\ell},a_{\ell}\right),\left(a_{\ell},b_1\right)$, where for every $j\in \{1,\ldots,\ell\}$, $(b_j,a_j)\in M_t$ and $(a_j,b_{j+1})\notin M_t$.
	
	For ease of notation, we define $\ell+1=1$. According to the definition of weights in $R_{>t}$, we know that $$\sum_{j=1}^{\ell}W_{>t}(\langle b_j,b_{j+1}\rangle)=\sum_{j=1}^{\ell}\left(v_{a_j}(b_j)-v_{a_j}(b_{j+1})\right)<0,$$ and therefore, $\sum_{j=1}^{\ell}v_{a_j}(b_{j+1})>\sum_{j=1}^{\ell}v_{a_j}(b_j)$. We get that the matching $M'$, which is constructed by removing the set $\{\left(b_j,a_j\right)\}_{j\in {1,\ldots,\ell}}$ from $M_{>t}$ and adding the set $\{\left(b_{j+1},a_j\right)\}_{j\in {1,\ldots,\ell}}$, is of larger weight, in contradiction to $M_{>t}$ being a maximum weight matching.

\end{Proof}

 We now process the relation graph $R_{>t}$:

 \begin{enumerate}
	\item Let $\Delta$ be the smallest total weight of a cycle with strictly positive total weight in $R_{>t}$, and let $\epsilon=\frac{\Delta}{\abs{\goods_{>t}}+1}$. Mark all edges in $E_{>t}$ that take part in \textbf{some} directed cycle of weight 0 in $R_{>t}$. Delete all marked edges. Let $E'_{>t}$ be the set of remaining edges. For every edge $e \in E'_{>t}$, set $W'_{>t}(e)=W_{>t}(e)-\epsilon$. Let $R'_{>t}=(\goods_{>t},E'_{>t},W'_{>t})$ be the resulting graph. \label{step:relation_graph}
	\item 
	Set $\price_{t+1}=\price$ where $\price$ is the output of $\fp$ (see Figure \ref{alg:find-assignment}) with $R'_{>t}$ as the input graph.
\end{enumerate}

\begin{figure}
	\MyFrame{
		
		$\fp$\\
		\textbf{Input:} A directed graph $G=(\goods,E,W)$ where all cycles are strictly positive.\\
		\textbf{Output:} a pricing function $\price:\goods\rightarrow \rep$ such that $p(b')-p(b)\geq -W(\langle b, b'\rangle)$ for every $\langle b, b'\rangle\in E$.
		\begin{enumerate}
			\item Add a dummy node $dum$, and draw an edge of weight $0$ from $dum$ to every other node.
			\item Compute the shortest path  from $dum$ to all nodes of $G$ (there are no negative cycles in $G$). For every $b\in\goods$, let $\dist_{dum}(b)$ denote the length of the shortest path from $dum$ to $b$.
			\item For every item $b\in \goods$, set $p(b) = -\dist_{dum}(b)$.
		\end{enumerate}
	} \caption{Pricing algorithm.}
	
	\label{alg:find-assignment}
\end{figure}

Since $\fp$ uses shortest paths computation, we need $R'_{>t}$ to have the following property.
\begin{lemma}
	All the directed cycles in ${R'}_{>t}$ are strictly positive. \label{cor:strictly_positive}
\end{lemma}
\begin{Proof}
	Let $\widetilde{R}$ be the graph which is obtained from $R_{>t}$ by removing all the edges that take part in a directed cycle of weight 0. Since according to Lemma \ref{lem:no_neg_cycles}, $R_{>t}$ has no negative weight cycles, all the cycles in $\widetilde{R}$ are of strictly positive weight. By the definition of $\Delta$, every simple cycle has a weight of at least $\Delta$. ${R'}_{>t}$ is constructed by taking $\widetilde{R}$ and decreasing all the edge weights by $\epsilon = \frac{\Delta}{\abs{\goods_{>t}}+1}$. Therefore, the weight of every simple cycle in $\widetilde{R}$ could have decreased by no more than $\abs{\goods_{>t}} \epsilon < \Delta$, which means that all the cycles in ${R'}_{>t}$ are of strictly positive weight.
\end{Proof}

\vspace{0.1in}

\setlength{\fboxrule}{2pt}
\noindent \fbox{\noindent\makebox[0.95\textwidth][c]{\begin{minipage}{0.9\textwidth}
	{\bf [Running example]} In Figure \ref{fig:all-phases}(b), the thick edges form a directed cycle of weight 0. We remove these edges and subtract $\epsilon$ from every remaining edge. We then run Algorithm $\fp$ on the obtained graph, which gives the prices presented in red next to each item in Figure \ref{fig:all-phases}(b). In this case, the only negative edge (after removing the cycle of length 0) is the edge $\langle d,a \rangle$, whose price is set to $-W'(\langle d,a \rangle)=-(-1-\epsilon)=1+\epsilon$. Since all the other shortest paths are positive, prices of other items do not change (recall the new price is the negation of the shortest path from the dummy item). When Alice arrives, she picks the unique item in her demand set --- item $b$. Similarly, graphs $G_{>t},R_{>t}$ of all iterations $t=0,1,2,3$ are demonstrated in Figure \ref{fig:all-phases}(c)-(h).
\end{minipage}}}\setlength{\fboxrule}{0.1pt} \par

\vspace{0.1in}

The following definition encompasses the properties needed by a pricing function in order to get the optimal social welfare by dynamic pricing.
\begin{defn}[{\good} prices1] \label{def:good_prices}
	Let $W_{>t}$ be a weight function as defined in Eq.~\eqref{eq:weight_def}, $R'_{>t}=(\goods_{>t},E'_{>t},W'_{>t})$ be a graph that's a result of step \ref{step:relation_graph} above. A pricing function $p:\goods_{>t}\rightarrow \rep$ is {\good} (equivalently, prices are {\good}) if the following three properties occur:
	\begin{eqnarray}
	\forall{b\in \goods_{>t}} & & \price(b)\geq 0 \label{lp:positive}\\
	\forall{\langle b_1,b_2\rangle\in E'_{>t}}  & & \price(b_1)-\price(b_2) < W_{>t}(\langle b_1,b_2\rangle) \label{lp:edge}\\
	\forall{b\in \goods_{>t}: \val_{>t}(b)>0} & & \price(b) < \val_{>t}(b) \label{lp:attractive} 	
	\end{eqnarray}
\end{defn}

Consider a directed edge $\langle b_1,b_2\rangle$ and some cycle it belongs to.  The edge $\langle b_1,b_2 \rangle$ came about because we choose a maximal matching where item $b_1$ was assigned to some buyer $a$, whereas $b_2$ was not. If all such cycles have strictly positive total weight, then  the edge weights, and the associated prices computed via $\fp$, ensure that agent $a$ prefers $b_1$ to $b_2$, effectively removing choices for ``wrong" tie breaking. Contrawise, if the edge $\langle b_1,b_2\rangle$ does belong to some cycle of total weight zero, this implies that the maximum matching is not unique. Ergo, whenever some item along this cycle is first chosen, it is still possible to extend the matching to a maximal weight matching. This is exactly where the dynamic pricing creeps in, subsequent to this symmetry breaking, new prices have to be computed to avoid wrong tie breaking decisions.

We now show that setting prices that satisfy the properties of {\good} prices ensures that after the arrival of all agents,
the social welfare achieved is maximized.

\begin{thm}
	A dynamic pricing scheme which calculates prices that are {\good} 
	achieves optimal social welfare (a maximum weight matching of $G$).\label{thm:matching_main}
\end{thm}
\begin{Proof}
Recall that $M_t$ is the matching which results from the first $t\in \{0,1,\ldots,\abs{\agents}\}$ agents taking an item which maximizes their utility and that $G_{>t}$ is the graph of the remaining agents and items after the first $t$ agents arrived and purchased some items. Let $M_{>t}$ be a maximum weight matching of $G_{>t}$, where $M_{>0}$ is a matching that maximizes the social welfare of all the agents, and $M_{>\abs{\agents}}=\emptyset$. We prove by induction that for every $t\in\{0,1,\ldots,\abs{\agents}\}$, $\SW(M_t)+\SW(M_{>t})=OPT$. It follows that the matching $M_{\abs{\agents}}$ yields optimal social welfare.

For $t=0$, this claim trivially holds since $\SW(M_{>0})=OPT$. Assume that for some $t-1$, $\SW(M_{t-1})+\SW(M_{>t-1})=OPT$. Let $\mpr$ be the maximum weight matching we compute at step $\ref{step:find_mathcing}.$ of the  pricing scheme. When agent $t$ arrives, consider the following cases:
\begin{itemize}
	\item Agent $t$ does not take any item. By property $(\ref{lp:attractive})$, the only case where an agent has no positive utility from any item is if she is matched to an item in $M_{t-1}$ with an edge of weight 0. In this case, $\SW(M_t)= \SW(M_{t-1})$, and by taking $M_{>t}$ to be the same matching as $M_{>t-1}$ without the edge the $t$-th agent is matched to, $\SW(M_{>t})=\SW(M_{>t-1})$. We get that $\SW(M_t)+\SW(M_{>t})=\SW(M_{t-1})+\SW(M_{>t-1})=OPT$.
	\item Agent $t$ takes the item which she is matched to in $M_{>t-1}$. Let $v$ be the value of the $t$-th agent for the item. Clearly, $\SW(M_t)=\SW(M_{t-1})+v$. By taking $M_{>t}$ to be the same matching as $M_{>t-1}$ without the edge the $t$-th agent is matched to,  we get $\SW(M_{>t})=\SW(M_{>t-1})-v$. We get that $\SW(M_t)+\SW(M_{>t})=\SW(M_{t-1})+v+\SW(M_{>t-1})-v=OPT$.
	\item Agent $t=a\in \agents_{>t-1}$ takes an item $b'\in \goods_{>t-1}$ which is different than $b\in \goods_{>t-1}$, the item which she is matched to in $M_{>t-1}$. Therefore,
	\begin{eqnarray}
	v_a(b')-\price_{t-1}(b')\geq v_a(b)-\price_{t-1}(b). \label{ineq:prices}
	\end{eqnarray}
	Let $\langle b,b'\rangle\in E_{>t-1}$ be the directed edge from $b$ to $b'$ in $R_{>t-1}$. Its weight $W_{>t-1}(\langle b,b'\rangle)= v_a(b)-v_a(b')$. If $\langle b,b'\rangle$ would have been in $R'_{>t-1}$, then according to property $(\ref{lp:edge})$, we would have had that $\price_{t-1}(b)-\price_{t-1}(b')< W_{>t-1}(\langle b,b'\rangle) = v_a(b)- v_a(b')$. Rearranging gives us $v_a(b')-\price_{t-1}(b') < v_a(b)-\price_{t-1}(b)$, which contradicts $(\ref{ineq:prices})$. Therefore, $\langle b,b'\rangle$ was removed from ${R'}_{>t-1}$, which can only happen if the edge is part of a directed cycle of weight 0 in $R_{>t-1}$.
	
	Let $b_1=b_{\ell+1}=b$, $b_2=b'$ and let $\langle b_1,b_2\rangle, \langle b_2,b_3\rangle,\ldots, \langle b_{\ell-1},b_{\ell}\rangle,\langle b_{\ell},b_{\ell+1}=b_{1}\rangle$ be a simple directed cycle of length $\ell$ and weight $0$ in $R_{>t-1}$ in which $\langle b,b'\rangle$ takes part. This cycle corresponds to a cycle of alternating edges in $G_{>t-1}$, $$\left(b_1=b,a_1=a\right)\left(a_1,b_2=b'\right),\left(b_2,a_2\right)\ldots\left(a_{\ell-1},b_{\ell}\right),\left(b_{\ell},a_{\ell}\right),\left(a_{\ell},b_{\ell+1}=b_1\right),$$ where
	$$(b_j,a_j)\in M_{>t-1}\mbox{\rm\ and\ }(a_j,b_{j+1})\notin M_{>t-1}\quad\mbox{\rm for every $j\in \{1,\ldots,\ell\}$}.$$ Since the directed cycle is of weight 0, we get that $$\sum_{j=1}^{\ell} W_{>t}(\langle b_j,b_{j+1}\rangle)=\sum_{j=1}^{\ell} \left(v_{a_j}(b_j)-v_{a_j}(b_{j+1})\right)=0,$$ which means that the value of the unmatched edges in the directed cycle, $\sum_{j=1}^{\ell} v_{a_j}(b_{j+1})$, is equal to the value of the matched edges, $\sum_{j=1}^{\ell} v_{a_j}(b_j)$.
	
	Let $\widetilde{M}_{>t-1}$ be the matching which is a result of taking $M_{>t-1}$, removing the edges in the set $\{(a_j,b_j)\}_{j\in \{1,\ldots,\ell\}}$, and adding the edges of $\{(b_{j+1},a_j)\}_{j\in \{1,\ldots,\ell\}}$; Note that $(a,b')=(a_1,b_2)\in\widetilde{M}_{>t-1}$. Since the edges we added to $\widetilde{M}_{>t-1}$ are of the same value as the edges we removed, $$\SW(\widetilde{M}_{>t-1})+\SW(M_{t-1})=\SW(M_{>t-1})+\SW(M_{t-1})=OPT.$$ We define $M_{>t}$ to be a matching comprised of the same edges as $\widetilde{M}_{>t-1}$ except $(a,b')$. Therefore, $\SW(M_{>t})=\SW(\widetilde{M}_{>t-1})-v_a(b')$. Clearly, we have that $\SW(M_t)=\SW(M_{t-1})+v_a(b')$. We get that $\SW(M_{>t})+\SW(M_t)=\SW(\widetilde{M}_{>t-1})-v_a(b')+\SW(M_{t-1})+v_a(b')=OPT$. This completes the proof of the induction and the theorem.
\end{itemize}	
\end{Proof}

It remains to show that $\fp$ output prices that are {\good} --- \textit{i.e.,} prices that satisfies all three properties in Definition \ref{def:good_prices}.
First, we observe that property $(\ref{lp:positive})$ is trivially satisfied. 
\begin{obs}
	$\fp$ computes prices which satisfy property $(\ref{lp:positive})$.\label{lem:type1}
\end{obs}
\begin{Proof}
	This follows since the length of the shortest path from $dum$ to every node is at most the length of the direct edge from $dum$ to this node, \textit{i.e.}, 0.
\end{Proof}

The following property is helpful in proving property $(\ref{lp:edge})$.
\begin{lemma}
	Let $G=(\goods,E,W)$ be the input graph of $\fp$ and let $\price:\goods\rightarrow \rep$ be its output. For every $\langle b_1, b_2\rangle\in E$ we have that $p(b_2)-p(b_1)\geq -W(\langle b_1, b_2\rangle)$.\label{lem:price_property}
\end{lemma}
\begin{Proof}
	Since the shortest path from $dum$ to $b_2$ is no longer than the shortest path from $dum$ to $b_1$ \textit{plus} the direct edge from $b_1$ to $b_2$, we have that $$\dist_{dum}(b_2)\leq \dist_{dum}(b_1)+W(\langle b_1,b_2\rangle).$$ Rearranging gives
	$$p(b_2)-p(b_1)=-\dist_{dum}(b_2)+\dist_{dum}(b_1)\geq -W(\langle b_1, b_2\rangle)$$ as desired.
\end{Proof}

We can now establish that property $(\ref{lp:edge})$ holds.

\begin{lemma}
	$\fp$ computes prices which satisfy property $(\ref{lp:edge})$.\label{lem:type2}
\end{lemma}
\begin{Proof}
		By Lemma \ref{lem:price_property}, we get that for a given $\langle b_1,b_2\rangle\in E'_{>t}$, $$\price(b_2)-\price(b_1)\geq -W'_{>t}	(\langle b_1,b_2\rangle)= -(W_{>t}(\langle b_1,b_2\rangle)-\epsilon).$$ Therefore, $$\price(b_1)-\price(b_2)\leq W_{>t}(\langle b_1,b_2\rangle)-\epsilon<W_{>t}(\langle b_1,b_2\rangle),$$ as desired.
\end{Proof}

For any two items $b,b'\in \goods$, let  $\dist(b,b')$ denote the length of the shortest path from $b$ to $b'$ in $R'_{>t}$.
For establishing that property $(\ref{lp:attractive})$ is met by the prices $\price(b)$'s computed by $\fp$, we need the following lemma.
\begin{lemma} \label{lem:sp_price}
	Let $b_{\ell}$ be some vertex with $\price(b_{\ell}) >0$, and let 
	$dum,b_0,b_1,\ldots,b_{\ell}$ be a shortest path from the dummy node $dum$ to $b_{\ell}$. For every $i\in \{0,1,\ldots,\ell\}$, $\price(b_i)=-\dist(b_0,b_i)$.
\end{lemma}
\begin{Proof}
	Let $b_i$ a vertex on the shortest path from $b_0$ to $b_{\ell}$. Since every sub-path of a shortest path is also a shortest path, it must be that $dum,b_0,\ldots,b_i$ is a shortest path from $dum$ to $b_i$, and that $\dist_{dum}(b_i)=W(\langle dum, b_0\rangle)+\dist(b_0,b_i) =\dist(b_0,b_i)$. Therefore, $\price(b_i)=-\dist_{dum}(b_i)=-\dist(b_0,b_i)$ as desired.
\end{Proof}

We get the the following two corollaries.
\begin{techcor}
	$\price(b_0)=0$. \label{cor:sp_source_price}
\end{techcor}

\begin{techcor}
	For every $i\in \{0,1,\ldots, \ell-1\}$, $\price(b_i)-\price(b_{i+1})=W_{>t}(\langle b_i,b_{i+1}\rangle)-\epsilon$.\label{cor:sp_diff}
\end{techcor}
\begin{Proof}
	Since every sub-path of a shortest path is also a shortest path, we get that $\dist(b_0,b_{i+1})=\dist(b_0,b_{i})+W'_{>t}(\langle b_{i},b_{i+1}\rangle)$. From Lemma \ref{lem:sp_price}, we get that $\price(b_i)= -\dist(b_0,b_{i})$ and
	\begin{eqnarray*}
		\price(b_{i+1}) & = & -\dist(b_0,b_{i+1})\\
		& = & -\dist(b_0,b_{i})-W'_{>t}(\langle b_{i},b_{i+1}\rangle)\\
		& = & \price(b_i)-(W_{>t}(\langle b_i,b_{i+1}\rangle)-\epsilon),
	\end{eqnarray*}
	where the last equality follows by the definition of $W'_{>t}$.	
\end{Proof}

We now prove that property $(\ref{lp:attractive})$ is met.

\begin{lemma}
	For every $b\in \goods_{>t}$ which is matched in $M_{>t}$ by a non-zero weight edge, $\price(b)< \val_{>t}(b)$. \label{lem:agent_preference}
\end{lemma}
\begin{Proof}
	Assume for the purpose of reaching a contradiction that there exists some $b=b_{\ell}$ which is matched in $M_t$ via an edge of strictly positive weight for which  $\price(b)\geq \val_{>t}(b)$.
	Let $dum,b_0,b_1,\ldots,b_{{\ell}}$ be a shortest path from $dum$ to $b_{\ell}$ in the graph processed in $\fp$. 
	According to Corollary \ref{cor:sp_diff}, for every $i\in \{0,1,\ldots,\ell-1\}$, $\price(b_i)-\price(b_{i+1})=W_{>t}(\langle b_i,b_{i+1}\rangle)-\epsilon$. Summing over all $i$'s gives us
	\begin{eqnarray}
	\sum_{i=0}^{\ell-1} W_{>t}(\langle b_i,b_{i+1}\rangle) = \price(b_0)-\price(b_{\ell}) + \ell\epsilon < -\price(b)+\Delta, \label{ineq1}
	\end{eqnarray}
	where the inequality stems from the fact that $\price(b_0)=0$ (Corollary \ref{cor:sp_source_price}), $b_\ell=b$, $\ell < \abs{\goods_{>t}}$ and $\epsilon=\frac{\Delta}{\abs{\goods_{>t}}+1}$.
	Let $a$ be the vertex that $b$ is matched to in $M_{>t}$. According to the definitions of the weights of edges in $R_{>t}$, we get that the  weight of the edge $\langle b,b_0\rangle\in E_t$ in $R_{>t}$ is
	\begin{eqnarray}
	W_{>t}(\langle b_{\ell},b_0\rangle)= v_{a}(b)-v_{a}(b_0)\leq \val_{>t}(b)\leq \price(b),\label{ineq2}
	\end{eqnarray}
	where the first inequality is due to the definition of $\val_{>t}(b)$, and the second inequality is due to our initial assumption. Combining $(\ref{ineq1})$ with $(\ref{ineq2})$ yields that the weight of the cycle \\$\langle b_0,b_1\rangle, \langle b_1,b_2\rangle,\ldots, \langle b_{\ell-1},b_{\ell}\rangle, \langle b_{\ell},b_{0}\rangle$ in $R_{>t}$ is $\sum_{i=0}^{\ell}W_{>t}(\langle b_i,b_{i+1\mod{\ell+1}}\rangle)< \Delta$. Since $\Delta$ is the minimal weight of a positive cycle in $R_{>t}$, we get that either the weight of the cycle is negative, which contradicts Lemma \ref{lem:no_neg_cycles}, or the cycle is of weight 0, contradicting the fact the we delete every edge that takes part in some cycle of weight 0 in $R_{>t}$ from ${R'}_{>t}$.
\end{Proof}

%% file: static_unique.tex
In the next section (Section \ref{sec:coverage}) we give an example where there is a unique optimum, there exist Walrasian prices over the items, and no dynamic bundle pricing scheme can guarantee an optimal outcome.

Here, we show that in the case of Gross Substitute valuations, a unique optimum implies the existence of {\sl static prices} that guarantee an optimal allocation (for any order of arrival).
Our pricing scheme is based on a combinatorial algorithm inspired by Murota \cite{murota1996valuated1,murota1996valuated2}.\footnote{See \cite{leme2014gross} for a concise description on how Murota's work relates to the computation of Walrasian prices for GS valuations.}

Given some set of items $A\subseteq \goods$, we define the sets of items local to $A$ as following $$\local(A)=\{B\neq A\subseteq \goods: \abs{B\setminus A}\leq 1\text{ and }\abs{A\setminus B}\leq 1 \}.$$ We present the following alternative definition of gross substitute valuations \cite{gul1999walrasian}:
\begin{defn}
	A valuation $v:2^\goods\rightarrow \rep$ satisfies the gross substitute condition whenever the following holds: for every item prices $\price:\goods \rightarrow \rep$, for every $A\subseteq \goods$ such that $A\notin \demand(\goods,\price)$ there exists $B\in \local(A)$ such that $\util(B,\price)> \util(A,\price)$.
\end{defn}
We refer to this characterization as the \textit{local improvement} property (LI).

A valuation function $v$ is submodular if for every $S,T$ such that $S\subseteq T \subseteq \goods$, and $j\in \goods$, $v(S\cup\{j\})-v(S)\geq v(T\cup\{j\})-v(T)$. We also use the characterization of Reijnierse et al. \cite{reijnierse2002verifying} for gross substitutes valuations:
\begin{defn}\label{def:gs3}
	A valuation $v:2^\goods\rightarrow \rep$ is gross substitutes if and only if $v$ is submodular, and for every $S\subset \goods$ and $b_1,b_2,b_3\notin S$:
	\begin{eqnarray}
	v(S\cup \{b_1,b_2\})+v(S\cup \{b_3\})\leq \max\{v(S\cup \{b_1\})+v(S\cup \{b_2,b_3\}),v(S\cup \{b_2\})+v(S\cup \{b_1,b_3\}) \}.\label{eq:gs_cond}
	\end{eqnarray}
\end{defn}

Given a set of gross-substitute valuations and items $\langle \vals, \goods\rangle$, let $\mathcal{B}=\{B_1,\ldots,B_n\}$ be the unique optimal allocation. We compute the prices $\price:\goods \rightarrow \rep$ as follows:
\begin{enumerate}
	\item Let $D=\{d_1,\ldots,d_n\}$ be a set of dummy items (one for each agent), $\goods'=\goods\cup D$ be the set of items after we added the dummy items. We extend every valuation function $v_i$ to the domain $2^{\goods'}$, where $v_i(X)=v_i(X\cap \goods)$ (i.e., the dummy items have no effect on the value of the bundle).
	Define $\mathcal{B'}=\{B'_1,\ldots,B'_n\}$ where every bundle $B'_i=\{B_i\cup \{d_i\}\}$ receives an additional dummy item.
	\item Let $R=\langle V=\goods', E\subset V\times V, W:E\rightarrow \Re\rangle$ (the exchange graph) be a weighted directed graph where:\label{step:ex_graph}
	\begin{itemize}
		\item $E=\{\langle a, b\rangle \in \goods'^2: a\in B'_i, b\in \goods' \setminus B'_i\text{ for every }i\}\setminus D^2$: I.e., there is an edge from every item in some bundle $B'_i$ to every item not in $B'_i$, \textit{unless} the two items are dummy items.
		\item Let $e=\langle a,b\rangle$ where $a\in B'_i$ of some agent $i$ be an edge in the graph. $W(e)=v_i(B'_i)-v_i(B'_i-a+b)$, i.e., the value of the agent from bundle $B'_i$ \textit{minus} the value she gets if she exchanges item $a$ for item $b$.
	\end{itemize}
	\item Let $\delta>0$ be the weight of a minimum weight cycle in $R$ (in the full version we show that
	all the cycles in $R$ are of strictly positive weight). Let $\gamma>0$ be the weight of the minimum weight path out of all the paths from any vertex to any dummy vertex (in the full version we show that all such paths are of strictly positive weight).
	Let $\epsilon=\frac{\min\{\delta,\gamma\}}{n+1}$.
	\item Update the weights by setting $W(e)\gets W(e)-\epsilon$ for every edge $e$ in the graph.\label{step:update_weight_unique}
	\item Price the items using algorithm $\fp$ (Figure \ref{alg:find-assignment}) with graph $R$ as input.
\end{enumerate}

We prove the following theorem\footnote{Independently, Paes Leme and Wong \cite{LemeW15} defined robust Walrasian pricing where there is no overlap between the demand sets of different buyers, and showed that for Gross substitute valuations with unique optima, such prices exist. Viewed from our perspective, this gives static prices that achieve optimal welfare for any order of arrival.}: 
\begin{thm}
\label{thm:unique-prices}
	Item prices $\price$ computed above achieve optimal welfare irrespective of the order of arrival of the buyers.
\end{thm}

Consider the algorithm for computing prices described above.
\begin{lemma}
	All the cycles in the graph $R$ described in step \ref{step:ex_graph} of the above price computation are of strictly positive weight.\label{lem:ex_grpah_pos}
\end{lemma}
\begin{Proof}
	Let $i$ be some agent (recall that $B_i$ the bundle allocated to her in the unique optimal allocation). Let $\delta = \min_{\allocs\neq \mathcal{B}}\{\SW(\mathcal{B},\vals)-\SW(\allocs,\vals)\}$ be the difference in welfare between the optimal allocation, and the second best allocation. $\delta > 0$ since the optimal allocation is unique. For some item $b\in \goods\setminus B_i$ define the modified valuation $\vmod:2^{\goods'}\rightarrow \rep$ as follows:
	\begin{eqnarray}
	\vmod(S)= \begin{cases}
	v_i(S)+\delta \quad & b\in S\\
	v_i(S)\quad & b\notin S
	\end{cases}.
	\end{eqnarray}
	Let $\valsmod=(v_{-i},\vmod)$. For some arbitrary allocation $\allocs\neq \bundles$ we have
	\begin{eqnarray*}
		\SW(\allocs,\valsmod)& = &\vmod(\alloci)+\sum_{j\neq i} v_j(\alloci[j])\\
		& \leq & \vali(\alloci)+\delta+\sum_{j\neq i} v_j(\alloci[j])\\
		& \leq & \SW(\bundles,\vals)\\
		& = & \SW(\bundles,\valsmod),
	\end{eqnarray*}
	and therefore, $\bundles$ is still an optimal allocation for profile $\valsmod$. We next claim that $\vmod$ is gross substitutes, by showing it satisfies the requirements in Definition~\ref{def:gs3}.
	
		First we show that $\vmod$ is submodular. Let $S\subset T$ two sets of items, and let $b'$ be some item. if $b'\neq b$, then we know that $\vmod(b' | S)=\vali(b'| S)\leq \vali(b'| T)= \vmod(b'|T)$. Otherwise, $\vmod(b' | S)=\vali(b'| S)+\delta\leq \vali(b'| T)+\delta= \vmod(b'|T)$. Next, we verify (\ref{eq:gs_cond}). Let $S$ be some set of items and $b_1,b_2,b_3$ some items not in $S$. Since, $v_i$ is GS, we know that (\ref{eq:gs_cond}) holds. Without loss of generality, let us assume that $v_i(S\cup \{b_1,b_2\})+v_i(S\cup \{b_3\})\leq v_i(S\cup \{b_1\})+v_i(S\cup \{b_2,b_3\})$, which is equivalent to $v_i(b_2|S\cup \{b_1\})\leq v_i(b_2|S\cup \{b_3\})$. If $b_2\neq b$ then $\vmod(b_2|S\cup \{b_1\})=v_i(b_2|S\cup \{b_1\})\leq v_i(b_2|S\cup \{b_3\})=\vmod(b_2|S\cup \{b_3\})$, and otherwise $\vmod(b_2|S\cup \{b_1\})=v_i(b_2|S\cup \{b_1\})+\delta\leq v_i(b_2|S\cup \{b_3\})+\delta=\vmod(b_2|S\cup \{b_3\})$. This implies that $\vmod(S\cup \{b_1,b_2\})+\vmod(S\cup \{b_3\})\leq \vmod(S\cup \{b_1\})+\vmod(S\cup \{b_2,b_3\})$.
		
		Since $\valsmod$ is a gross substitute valuation profile, it admits a Walrasian equilibrium $(\bundles',\price)$. We claim that $(\bundles',\price)$ is also a Walrasian equilibrium for $\vals$. This is true since $\vali(B'_i)=\vmod(B'_i)$, and for every $S$, $\vali(S)\leq\vmod(S)$.
		
		For some item $b'\in \goods'$, we denote by $\agents(b')$ the function that returns the agent $j$ for which $b'\in B'_j$. Consider a cycle in $R$ that uses edge $\langle a, b\rangle$ for some cycle in $R$. Let $(b_0,b_1,\ldots , b_{\ell-1}, b_0)$ denote the cycle, where $b_0=a$ and $b_1=b$. We denote $b_{\ell}=b_0$. Since $(\bundles', \price)$ is a Walrasian equilibrium for $\valsmod$, we know that
		\begin{eqnarray*}
			\vali(B'_i)-\price(B'_i) & = &\vmod(B'_i)-\price(B'_i)\\
			& \geq & \vmod(B'_i-a+b)-\price(B'_i-a+b)\\
			& = & \vali(B'_i-a+b) + \delta -\price(B'_i-a+b) \\
			& > & \vali(B'_i-a+b) -\price(B'_i-a+b).
		\end{eqnarray*}
		Rearranging gives us
		\begin{eqnarray}
			W(\langle b_0,b_1\rangle) & = & W(\langle a,b\rangle)\nonumber\\
			& = & \vali(B'_i)-\vali(B'_i-a+b)\nonumber\\
			& > & \price(B'_i)-\price(B'_i-a+b)\nonumber\\
			& = & \price(a)-\price(b)\nonumber\\
			& = & \price(b_0)-\price(b_1).\label{eq:strict_ineq}
		\end{eqnarray}
		Since $(\bundles', \price)$ is a Walrasian equilibrium for $\vals$ as well, we get that for every $j\in {1,\ldots, \ell-1}$, $$\vali[\agents(b_j)](B'_{\agents(b_j)})-\price(B'_{\agents(b_j)})\geq \vali[\agents(b_j)](B'_{\agents(b_j)}-b_j+b_{j+1})-\price(B'_{\agents(b_j)}-b_j+b_{j+1}).$$
		Rearranging gives us
		\begin{eqnarray}
		W(\langle b_j,b_{j+1}\rangle) & = & \vali[\agents(b_j)](B'_{\agents(b_j)})-\vali[\agents(b_j)](B'_{\agents(b_j)}-b_j+b_{j+1})\nonumber\\
		& \geq & \price(b_j)-\price(b_{j+1}).\label{eq:wal_ineq}
		\end{eqnarray}
		Summing inequality (\ref{eq:strict_ineq}) with inequalities of type (\ref{eq:wal_ineq}) for all $j\in {1,\ldots, \ell-1}$ gives us that the weight of the cycle $(b_0,b_1,\ldots , b_{\ell-1}, b_0)$ is $$\sum\limits_{j\in\{0,\ldots, \ell-1\}} W(\langle b_j,b_{j+1}\rangle) > \sum\limits_{j\in\{0,\ldots, \ell-1\}}\left(\price(b_j)-\price(b_{j+1})\right)=0.$$
		Since agent $i$ is an arbitrary agent and item $b$ is an arbitrary (non-dummy) item, we get that all the cycles in $R$ that use an edge which ends in a non-dummy item must be strictly positive. Since there are no edges who between two dummy items in $R$, we get that all cycles must use at least one edge which ends in a non-dummy item, hence, must be strictly positive.
\end{Proof}

We now show a property which is crucial in establishing that the price of every dummy node is zero.
\begin{lemma}
	Let $R$ be the graph described in step \ref{step:ex_graph} of the above price computation. For every agent $i$, dummy node $d_i$ and every item $b\in \goods'\setminus\{d_i\}$, $\dist_R(b,d_i)>0$.\label{lem:pos_paths}
\end{lemma}
\begin{Proof}
	Let $d_i$ be a dummy item added to the bundle of some agent $i$. Let $b$ be some item in $\goods'\setminus \{d_i\}$.
	For some dummy item $d_j\neq d_i$, let $R_{d_i,d_j}$ be the graph established by taking graph $R$ (after step \ref{step:ex_graph}), and adding an edge $\langle d_ i, d_j\rangle$ of weight $W(\langle d_i, d_j\rangle)=V(B'_i)-V(B'_i-d_i+d_j)=0$.
	First notice using a similar argument to the one presented in the proof of Lemma \ref{lem:ex_grpah_pos}, it is not hard to see that all the cycles in the graph $R_{d_i,d_j}$ are of strictly positive weight for any choice of $d_j$. We use $b\rightsquigarrow d_i$ and $W(b\rightsquigarrow d_i)$ to denote some simple path from $b$ to $d_i$ and its weight.
	We now consider the following cases:
	\begin{itemize}
		\item $b$ is in $\goods\setminus B_i$: In this case, consider the cycle obtained by adding edge $\langle d_i,b\rangle$ to $b\rightsquigarrow d_i$. Since every cycle in $R$ is of strictly positive weight, we have that $W(b\rightsquigarrow d_i)+W(\langle d_i, b\rangle)>0$. Since $W(\langle d_i, b\rangle)=v_i(B_i)-v_i(B_i+b)\leq 0$, it must be the case where $W(b\rightsquigarrow d_i)>0$.
		\item $b$ is some dummy item $d_j\neq d_i$: Consider the graph $R_{d_i,d_j}$ and the cycle obtained by adding edge $\langle d_i, d_j\rangle$ to $d_j\rightsquigarrow d_i$. Since every cycle in $R_{d_i,d_j}$ is of strictly positive weight, we have $W(d_j\rightsquigarrow d_i)+W(\langle d_i, d_j\rangle)= W(d_j\rightsquigarrow d_i)>0$.
		\item $b\in B_i$: Consider the graph $R_{d_i,d_j}$. Consider the cycle obtained by adding edges $\langle d_i, d_j\rangle,\langle d_j, b\rangle$ to $d_j\rightsquigarrow d_i$. We have that the weight of the cycle is $$W(b\rightsquigarrow d_i)+W(\langle d_i, d_j\rangle)+W(\langle d_j, b\rangle)= W(d_j\rightsquigarrow d_i)+W(\langle d_j, b\rangle)>0.$$
		Since $W(\langle d_j, b\rangle)=v_j(B_j)-v_j(B_j+b)\leq 0$, we get $W(b\rightsquigarrow d_i)>0$.
	\end{itemize}
	Since $W(b\rightsquigarrow d_i)>0$ for every simple path from $b$ to $d_i$, and there are no negative cycles in $R$, we have that $\dist_R(b,d_i)>0$.
	
\end{Proof}

From Lemmas \ref{lem:ex_grpah_pos} and \ref{lem:pos_paths} and by carefully choosing $\epsilon$ in step \ref{step:update_weight_unique}, we immediately get:
\begin{techcor}
	 \textbf{After} updating the edge weights (step \ref{step:update_weight_unique}) all the cycles in $R$ are of \textbf{strictly} positive weight, all the paths ending in a dummy vertex are of a \textbf{strictly} positive weight. \label{cor:still_pos}
\end{techcor}

It is crucial that we have the following:
\begin{techcor}
	For every dummy item $d_i$, $\price(d_i)=0$.
\end{techcor}
\begin{Proof}
	By the way $\fp$ operates, an item $d_i$ has a price greater than 0 only if there exists a path of negative weight from some vertex to $d_i$. By Corollary \ref{cor:still_pos}, this cannot happen.
\end{Proof}

The next lemma shows that for every ``local" change an agent may perform, her utility decreases.

\begin{lemma}
	For every agent $i$, for every bundle $C\in \local(B_i)$, we have $\util(\price,B_i)>\util(\price,C)$. \label{lem:local_disimprov}
\end{lemma}
\begin{Proof}
	Let $C$ be some bundle in $\local(B_i)$. We inspect the following cases:
	\begin{itemize}
		\item $A\setminus C=\{a\}$ and $C\setminus A=\{b\}$: In this case, there is a directed edge in $\langle a,b\rangle\in E$ of weight $v_i(B_i)-v_i(B_i-a+b)-\epsilon=v_i(B_i)-v_i(C)-\epsilon$. By Lemma \ref{lem:price_property}, $\price(C)- \price(B_i)\geq-v_i(B_i)+v_i(C)+\epsilon>v_i(C)-v_i(B_i).$
		Rearranging gives us $\util(\price,B_i)=v_i(B_i)-\price(B_i)>v_i(C)-\price(C)=\util(\price,B_i)$.
		\item $A\setminus C=\{a\}$ and $C\setminus A=\emptyset$: There is an edge between $a$ and some dummy item $d_j$ of weight $v_i(B_i)-v_i(B_i-a+d_j)-\epsilon=v_i(B_i)-v_i(B_i-a)-\epsilon=v_i(B_i)-v_i(C)-\epsilon$. Again, by Lemma \ref{lem:price_property} we get that  $\util(\price,B_i)>\util(\price,B_i)$.
		\item $A\setminus C=\emptyset$ and $C\setminus A=\{b\}$: There is an edge between $d_i$ and $b$ of weight $v_i(B_i)-v_i(B_i-d_i+b)-\epsilon=v_i(B_i)-v_i(B_i+b)-\epsilon=v_i(B_i)-v_i(C)-\epsilon$. Again, by Lemma \ref{lem:price_property} we get that  $\util(\price,B_i)>\util(\price,B_i)$.
	\end{itemize}	
\end{Proof}

The following property of gross substitute valuations shows that the above lemma is enough to show that the prices achieve optimal social welfare.
\begin{lemma}
	Let $v:\goods \rightarrow \rep$ be a valuation that satisfies the gross substitute property, let $P:\goods\rightarrow \rep$ be some item pricing and let $A$ be some set of items in $\demand(\goods,\price)$. If $\abs{\demand(\goods,\price)}>1$ then there exists some $B\in \local(A)$ such that $B\in \demand(\goods,\price)$.\label{lem:unique_demand}
\end{lemma}
\begin{Proof}
	Let $A$ be some set in $\demand(\goods,\price)$ and let us assume that $\abs{\demand(\goods,\price)}>1$ and $\demand(\goods,\price)\cap \local(A)=\emptyset$. Let us define the following set:
	\begin{eqnarray*}
		\local^+(\price,A)=\{B\in \local(A): \exists C\neq A\in \demand(\goods,\price)\text{ s.t. } \abs{B\Delta C}\leq \abs{A\Delta C}\},
	\end{eqnarray*}
	that is, the set of local sets to $A$ that are more similar to another set in $\demand(\goods,\price)$ than $A$ is. Since $\abs{\demand(\goods,\price)}>1$, $\local^+(\price,A)$ is non empty. Let $B=\arg\min_{X\in \local^+(\price,A)}\{u(\price,A)-u(\price,X)\}$, let $C\neq A$ be the set in $\demand(\goods,\price)$ such that $\abs{B\Delta C}\leq \abs{A\Delta C}$ and let $\delta=\min_{X\in \local(A)}\{u(\price,A)-u(\price,X)\}$. $\delta>0$ by our assumption. We define the following item pricing $\price'$:
	\begin{itemize}
		\item If $\abs{B\setminus A}=1$, then for $a\in B\setminus A$ set $\price'(a)=\price(a)-\delta/2$ and $\price'(b)=\price(b)$ for all other $b\in \goods-a$.
		\item Otherwise, let $a$ be an item in $A\setminus B$. Set $\price'(a)=\price(a)+\delta/2$ and $\price'(b)=\price(b)$ for all other $b\in \goods-a$.
	\end{itemize}
	Notice that $C\in \demand(\goods,\price')$, $A\notin \demand(\goods,\price')$ and $\demand(\goods,\price')\subset \demand(\goods,\price)$. Therefore, $\local^+(\price',A)\subseteq \local^+(\price,A)$. If $\abs{B\setminus A}=1$ then for every set $X\in \local(A)$, we have that $\util(\price',A)= \util(\price,A)\geq \util(\price,X)+\delta > \util(\price',X)$. Otherwise, for every $X\in \local(A)$, $\util(\price',A)= \util(\price,A) - \delta/2 \geq \util(\price,X)+\delta/2 > \util(\price',X)$. Therefore, $A\notin \D(\goods,\price)$, and there is no local improvement, contradicting the LI property of gross substitute valuations.
\end{Proof}

To conclude the proof of Theorem \ref{thm:unique-prices}, we observe that by Lemma \ref{lem:local_disimprov} for every agent $i$ $u_i(\price,B_i)>  u_i(\price,X)$ for every $X\in \local(B_i)$. By the LI property of $v_i$, we have that $B_i\in \demandi(\goods,\price)$. By Lemma \ref{lem:unique_demand} we get that $\demandi(\goods,\price)=\{B_i\}$.

%% file: coverage.tex
We show an instance with agents with coverage valuations for which no dynamic pricing scheme guarantees an optimal allocation. Interestingly, this instance admits Walrasian prices and has a unique optimal allocation, so no combination of these conditions is sufficient to imply optimal dynamic pricing schemes. 

\begin{thm}
There exists an instance with agents with coverage valuations such that no dynamic pricing scheme guarantees more than a fraction $\frac{7.5}{8}$ of the optimal social welfare. This instance admits Walrasian prices.\label{thm:coverage_main}
\end{thm}
\begin{Proof}
 Let $\goods=\{a,b,c,d\}$ be a set of items and $\agents=\{1,2,3,4\}$ be a set of agents. Agents $2,3$ and $4$ are unit demand with the following valuation functions:
 \begin{eqnarray*}
 	v_2(S)=\begin{cases}
 		2\quad& S\cap \{a,b\}\neq \emptyset\\
 		0\quad  &otherwise
 	\end{cases}, v_3(S)=\begin{cases}
 	2\quad& S\cap \{a,c\}\neq \emptyset\\
 	0\quad  &otherwise
 \end{cases}, v_4(S)=\begin{cases}
 1\quad& S\cap\{d\}\neq \emptyset\\
 0\quad  &otherwise
\end{cases}.
\end{eqnarray*}

In addition, agent $1$ has the following coverage valuation:
\begin{eqnarray*}
	v_1(S)=\begin{cases}
		2\quad& S= \{b\},S=\{c\}\\
		3\quad& S= \{a\},S=\{d\}\\
		3.5\quad& S=\{a,b\},S=\{a,c\},S=\{d,b\},S=\{d,c\},S=\{a,d\}\\
		3.75\quad& S=\{a,b,d\},S=\{a,c,d\}\\
		4\quad  & \{b,c\} \subseteq S
	\end{cases}.
\end{eqnarray*}

Coverage valuation: To see that this is a coverage valuation, consider the following explicit representation. Let $\{e_1,e_2,e_3,e_4,e_5,e_6,e_7,e_8\}$ be the set of elements, with weights $w(e_1)=w(e_5)=5/4$ and $w(e_i)=1/4$ for $i \neq 1,5$. Item $a$ covers the set $\{e_1,e_2,e_5,e_6\}$, item $b$ covers the set $\{e_1,e_2,e_3,e_4\}$, item $c$ covers the set $\{e_5,e_6,e_7,e_8\}$, and item $d$ covers the set $\{e_1,e_4,e_5,e_8\}$.

Unique optimal allocation: The unique optimal allocation is to allocate item $a$ to agent 1, item $b$ to agent 2, item $c$ to agent 3 and item $d$ to agent 4. This allocation obtains social welfare of 8.

Walrasian prices: One can easily verify that the unique optimal allocation along with pricing each item at 1 is a Walrasian equilibrium.

We now show that no dynamic pricing scheme guarantees more than a fraction $\frac{7.5}{8}$ of the optimal allocation. In order to guarantee an optimal allocation, the following conditions must be satisfied:
\begin{itemize}
	\setlength{\itemsep}{1pt}
	\setlength{\parskip}{0pt}
	\setlength{\parsep}{0pt}
	\item Agent 4's utility from item $d$ should be strictly positive; i.e.,
	\begin{eqnarray}
	\price(d)<v_4(d)=1. \label{eq:cond1}
	\end{eqnarray}
	\item Agent 1 should strictly prefer item $a$ over item $d$, i.e.,
	\begin{eqnarray}
	v_1(a)-\price(a)>v_1(d)-\price(d)\Rightarrow \price(a)< \price(d). \label{eq:cond2}
	\end{eqnarray}
	\item Agent 2 should strictly prefer item $b$ over item $a$, i.e.,
	\begin{eqnarray}
	v_2(b)-\price(b)>v_2(a)-\price(a)\Rightarrow \price(b)< \price(a). \label{eq:cond3}
	\end{eqnarray}
	\item Agent $3$ should strictly prefer item $c$ over item $a$, i.e.,
	\begin{eqnarray}
	\price(c)< \price(a). \label{eq:cond4}
	\end{eqnarray}
	\item Agent 1 should strictly prefer item $a$ over the bundle $\{b,c\}$, i.e.,
	\begin{eqnarray}
	v_1(a)-\price(a)>v_1(\{b,c\})-\price(b)-\price(c)\Rightarrow \price(b)+\price(c)-\price(a)>1. \label{eq:cond5}
	\end{eqnarray}
\end{itemize}
Combining Equations (\ref{eq:cond1}) and (\ref{eq:cond2}) implies that $\price(a)<1$, while combining Equations (\ref{eq:cond3}), (\ref{eq:cond4}) and (\ref{eq:cond5}) yields $\price(a)>1$. Therefore, for every prices one might set, the adversary can set an order for which the first agent picks a different item than the one allocated to her in the optimal allocation.

Remark: note that the valuation function of agent $1$ is not gross substitutes. In particular, her demand under prices $\price(a)=\price(d)=\epsilon$, $\price(b)=\frac{1}{4}+\epsilon$ and $\price(c)=0$ is $\{b,c\}$, but if the price of item $c$ increases to $2$, then the unique bundle in the demand of agent $1$ is $\{a,d\}$.
\end{Proof} 

%% file: static_half.tex
In this section we show that, given a partition of the items into bundles, pricing each bundle half of its value to the buyer guarantees half of the social welfare of the partition. Let $\bundles=\{\bundle_1,\bundle_2,\ldots,\bundle_n\}\in \left(2^\goods\right)^n$ be a partition of the items such that $\bigcup_i \bundle_i=\goods$ and for every $i\neq j$ $\bundle_i \cap \bundle_j=\emptyset$. Let $W=\sum_i v_i(\bundle_i)$. We have the following:
\begin{thm}
	Let $\price:\bundles\rightarrow\rep$ be static bundle prices such that for every $i$, $\price(\bundle_i)=v_i(\bundle_i)/2$. This pricing scheme achieves a welfare of at least $W/2$.\label{thm:half_main}
\end{thm}
\begin{Proof}
	Let $\allocs$  be an allocation which is a result of agents arriving at an arbitrary order, each taking their favorite bundles.
	Notice that the utility of an agent for acquiring the bundles in $\alloc_i$ is $u_i(\alloc_i, P)=v_i(\bigcup_{B\in \alloc_i} B)-\sum_{B\in \alloc_i} \price(B)$. Let $\mathbb{I}_i$ be an indicator which gets 1 if bundle $\bundle_i$ was acquired by some agent and 0 otherwise. Rearranging and summing over all the agents gives us:
	\begin{eqnarray}
	\sum_i v_i\left(\bigcup_{B\in \alloc_i} B\right) & = & \sum_i \left(u_i(\alloc_i, P) + \sum_{B\in \alloc_i} \price(B)\right) \nonumber\\
	& = & \sum_i u_i(\alloc_i, P) + \mathbb{I}_i \price(\bundle_i).\label{eq:val_eq}
	\end{eqnarray}
	We show that for every $i$, $u_i(\alloc_i, P) + \mathbb{I}_i \price(\bundle_i)\geq v_i(\bundle_i)/2$. Using $(\ref{eq:val_eq})$ this is enough to prove the claim. For some $i$, either bundle $B_i$ is purchased by some agent, in which case $\mathbb{I}_i \price(\bundle_i)=v_i(\bundle_i)/2$. Otherwise, when agent $i$ arrived, she could have purchased bundle $\bundle_i$, for which she would have gotten a utility of $v_i(\bundle_i)-\price(\bundle_i)=v_i(\bundle_i)/2$. Since she bought the bundles which maximized her utility, her utility can only be greater than that, meaning $u_i(\alloc_i, \price)\geq  v_i(\bundle_i)/2$.
\end{Proof}

%% file: superadd_new.tex
We show that in the case where all agents have super-additive valuations, it is possible to come up with bundles and bundle-prices such that for every arrival order of the agents, the resulting welfare is optimal. 
The pricing algorithms takes a set of bundles $\bundles$ and an optimal allocation of bundles to agents $\allocs:\agents\rightarrow \bundles \cup \{\emptyset\}$ ($\alloci\cap \alloc_j=\emptyset$) such that the resulting welfare of the allocation is $\sum_i v_i(\alloci)=\OPT$, and outputs a possibly different allocation $\bundles'$ and supporting prices $\price:\bundles'\rightarrow \rep$ that ensure an optimal welfare regardless of agents' order and tie breaking.

\begin{figure}[H]
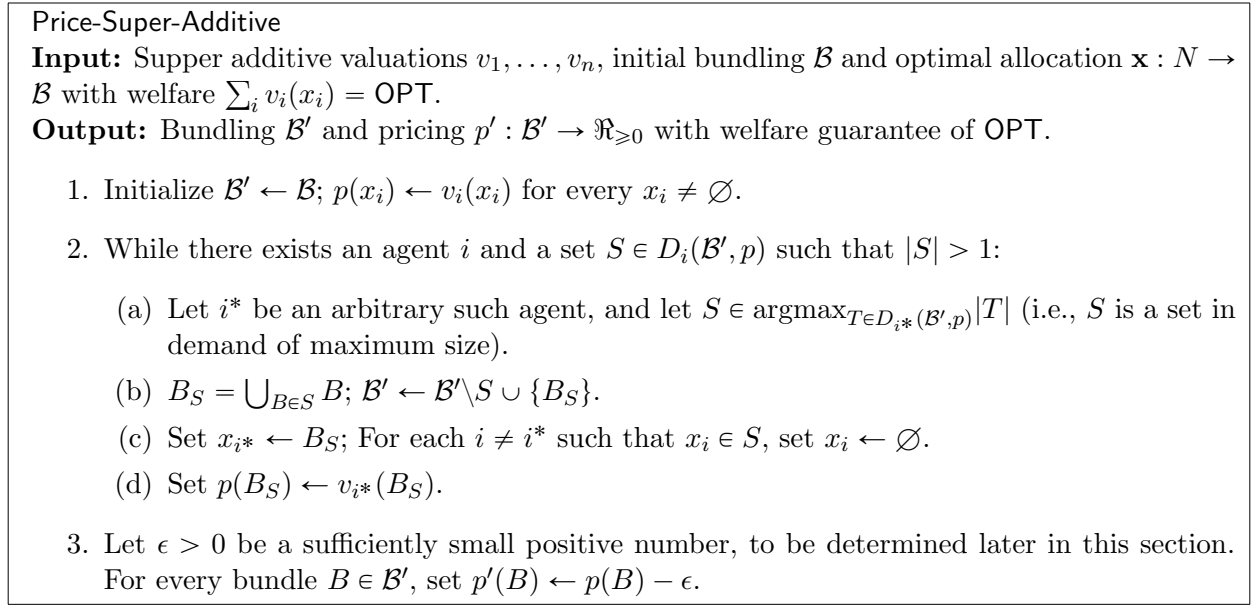

	\MyFrame{
		
		$\sapb$\\
		\textbf{Input:} Supper additive valuations $v_1,\ldots , v_n$, initial bundling $\bundles$ and optimal allocation $\allocs:\agents\rightarrow\bundles$  with welfare $\sum_i \vali(\alloci)=\OPT$.\\
		\textbf{Output:} Bundling $\bundles'$ and pricing $\price':\bundles'\rightarrow \rep$ with welfare guarantee of $\OPT$.
		\begin{enumerate}
			\item Initialize $\bundles'\gets \bundles$; $\price(\alloci)\gets v_i(\alloci)$ for every $\alloci\neq \emptyset$.
			\item While there exists an agent $i$ and a set $S \in D_{i}(\mathcal{B'},\price)$ such that $|S|>1$: \label{step:merge_bundles}
			\begin{enumerate}
				\item Let $i^*$ be an arbitrary such agent, and let $S \in \argmax_{T \in D_{i^*}(\mathcal{B'},\price)}|T|$ (i.e., $S$ is a set in demand of maximum size).\label{step:is_demand}
				\item $\bundle_S=\bigcup_{\bundle\in S} \bundle$; $\bundles'\gets \bundles'\setminus S\cup \{\bundle_S\}$.
				\item Set $\alloc_{i^*}\gets \bundle_S$; For each $i\neq i^*$ such that $\alloci\in S$, set $\alloci\gets \emptyset$.
				\item Set $\price(\bundle_S)\gets v_{i^*}(\bundle_S)$.				
			\end{enumerate}
			\item Let $\epsilon>0$ be a sufficiently small positive number, to be determined later in this section. For every bundle $\bundle\in\bundles'$, set $\price'(\bundle)\gets \price(\bundle) - \epsilon$. \label{step:decrease_prices}
		\end{enumerate}
	} \caption{Computing bundle prices for super-additive valuations.}
	
	\label{alg:supadd-bundle-prices}
\end{figure}

Recall that $\demandi(\bundles, \prices)$ returns all the sets of bundles that maximize agent $i$'s utility at the given prices.
Consider the process given in Figure \ref{alg:supadd-bundle-prices}.

The above process is guaranteed to terminate, since in every iteration of Step \ref{step:merge_bundles}, the number of bundles is decreased by at least~1. The next lemma ensures that after every iteration, all items are still allocated.
\begin{lemma}
	Consider the set $S$ defined in Step \ref{step:is_demand} of the above process, and the allocated bundle to $i^*$, $\alloc_{i^*}$, at this point. If $\alloc_{i^*}\neq \emptyset$, then $\alloc_{i^*}\in S$.
\end{lemma}
\begin{proof}
	Assume this is not the case. By super additivity, $v_{i^*}(S\cup \{\alloc_{i^*}\})\geq v_{i^*}(S)+v_{i^*}(\alloc_{i^*}).$ Since at this point, $\price(\alloc_{i^*})$ is exactly $v_{i^*}(\alloc_{i^*})$, the utility of agent $i^*$ for $S\cup \{\alloc_{i^*}\}$ is
	\begin{eqnarray*}
		v_{i^*}(S\cup \{\alloc_{i^*}\}) - \sum_{\bundle\in S} \price(\bundle) - \price(\alloc_{i^*}) \geq v_{i^*}(S)+v_{i^*}(\alloc_{i^*})- \sum_{\bundle\in S} \price(\bundle) - v_{i^*}(\alloc_{i^*}) = v_{i^*}(S)- \sum_{\bundle\in S} \price(\bundle).
	\end{eqnarray*}
	The term in the right hand side is exactly the utility of agent $i^*$ from the set of bundles $S$ at this point. This implies that $S\cup\{\alloc_{i^*}\}$ grants agent $i^*$ at least as much utility as $S$, contradicting the fact that $S$ was the largest set of bundles in the demand of $i^*$ at this point.
\end{proof}

Next, we show that the social welfare never decreases during the process.
\begin{lemma} \label{lem:opt_welfare}
	After each iteration of Step \ref{step:merge_bundles}, $\sum_i v_i(x_i)= \OPT$.
\end{lemma}
\begin{proof}
	Let $i^*$ and $S$ be the agent and set of bundles defined in Steps \ref{step:merge_bundles} and \ref{step:is_demand} of the above process. Let $\agents_S=\{i\in \agents: x_i\in S\}$. In order to prove the assertion of the lemma, we show that $v_{i^*}(S) \geq \sum_{i\in \agents_S} v_i(x_i)$. Since the utility of each set of bundles in the demand is non-negative (the emptyset is always feasible), it holds that
	$$v_{i^*}(S)-\sum_{\bundle\in S} \price(\bundle) = v_{i^*}(S)-\sum_{i\in \agents_S} \price(\alloci) = v_{i^*}(S)-\sum_{i\in \agents_S} v_i(\alloci)\geq 0.$$
	By rearranging, we get $v_{i^*}(S)\geq \sum_{i\in \agents_S} v_i(\alloci)$. 
That is, the welfare has not decreased by the reallocation of bundles in $S$.
\end{proof}

\begin{lemma}\label{lem:demand_structure}
	At the end of Step \ref{step:merge_bundles}
	\begin{itemize}
		\item For each agent $i$ such that $\alloci\neq \emptyset$, the only non-empty set of bundles in the demand of agent $i$ is the singleton bundle $\{\alloci\}$, for which the utility is~0.
		\item For each agent $i$ such that $\alloci=\emptyset$, there is at most one non-empty set of bundles in the demand of this agent. In the case there is one, it is a singleton bundle for which the utility is~0.
	\end{itemize}
\end{lemma}
\begin{proof}
	Consider an agent $i$ at the end of Step \ref{step:merge_bundles}. We first notice that every non-empty set of bundles in the demand of the agent must be a singleton bundle. Otherwise, by loop's condition in Step \ref{step:merge_bundles}, this process would not have terminated.
	For every agent $i$ such that $\alloci\neq\emptyset$, since we set $\pricei(\alloci)$ to be exactly $v_i(\alloci)$, the utility of the agent for $\alloci$ is~0. For agent $i$ with $\alloci=\emptyset$, if $i$ has a strictly positive utility for some bundle $\alloc_j$ of agent $j\neq i$, then $v_i(\alloc_j)>\price(\alloc_j)=v_j(\alloc_j)$, and the allocation is not optimal, in contradiction to Lemma \ref{lem:opt_welfare}. Therefore, in this case, the utility for any bundle in the demand of agent $i$ is~0 as well.
	
	Next, we notice there cannot be two singletons $\{B_1\}, \{B_2\}$ in the demand of agent $i$, since in this case $v_i(B_1) \geq \price(B_1)$, and therefore, the utility of agent $i$ for $\{B_1,B_2\}$ is at least $$v_i(\{B_1,B_2\})-\price(B_1)-\price(B_2)\geq v_i(B_1)+v_i(B_2) -\price(B_1)-\price(B_2)\geq v_i(B_2) -\price(B_2),$$ where the first inequality follows from sub-additivity. This implies that the set $\{B_1,B_2\}$, must be in the demand of agent $i$ as well, which contradicts the fact that all sets in the demand are singleton bundles.
	
	Lastly, we notice that for agent $i$ such that $\alloci\neq \emptyset$, if there was a set $\{B\}$ in the demand and $B\neq \alloci$, then since $\price(\alloci)=v_i(\alloci)$, the utility of $i$ for set $\{B,\alloci\}$ is
	$$v_i(\{B,\alloci\})-\price(B)-\price(\alloci)\geq v_i(B)+v_i(\alloci) -\price(B)-\price(\alloci)= v_i(B)-\price(B),$$ implying that $\{B,\alloci\}$ is in the demand of agent $i$ as well, a contradiction. This completes the proof of the lemma.
\end{proof}

Consider some agent $i$ at the end of Step \ref{step:merge_bundles} of $\sapb$. Since the utility of agent $i$ for the singleton bundle in the demand (if there is one) is~0 according to the above lemma, agent $i$ has negative utility for every other non-empty set of bundles.
Therefore, for each agent $i$, we define $$\delta_i=\min_{\{S\subseteq \bundles':S\notin \demandi(\bundles', \prices)\}} -\utili(S,\price),$$ where $\prices$ are the prices at the end of Step \ref{step:merge_bundles}. Roughly speaking, this indicates by how much the utility of the agent for any set not in the demand can increase while keeping this set not in the demand. Define the following $\epsilon$, to be used to decrease the price of every bundle in $\bundles'$ as described in Step \ref{step:decrease_prices} of the above process.
\begin{eqnarray}
	\epsilon=\frac{\min_i \delta_i}{n+1}. \label{eq:eps_definition}
\end{eqnarray}
The following theorem states that if we set prices by using $\sapb$, then the resulting welfare is optimal.
\begin{thm}\label{thm:superadditive_main}
	For any setting with super-additive valuations, there exist bundling and static prices over bundles such that for any arrival order of the agents (and any tie breaking of the agents), the resulting allocation is optimal.
\end{thm}
\begin{proof}
	Consider some set of bundles $S\subseteq \bundles'$ which is not in the demand of agent $i$ at the end of Step $2$ of $\sapb$. By the definition of $\delta_i$, at the end of Step \ref{step:merge_bundles}, $\sum_{\bundle\in S} \price(\bundle)- v_i(S)\geq \delta_i.$ Since at Step \ref{step:decrease_prices}, each bundle's price is decreased by $\epsilon$ (as defined in \eqref{eq:eps_definition}), we have that after Step \ref{step:decrease_prices},
	\begin{eqnarray*}
		\sum_{\bundle\in S} \price'(\bundle)- v_i(S)& \geq & \delta_i - |S|\cdot \epsilon\\
		&= & \delta_i-|S|\cdot \frac{\min_i \delta_i}{n+1}\\
		&\geq & \delta_i-\frac{|S|\cdot \delta_i}{n+1}\\
		& > &\delta_i - \delta_i = 0,
	\end{eqnarray*}
	where the last inequality follows because there are at most $n$ bundles in $\bundles'$ (and therefore in $S$). Therefore, the utility of agent $i$ for every set of bundles not in his demand at the end of Step \ref{step:merge_bundles} is negative, while $i$'s utility for the singleton bundle in $i$'s demand (if exists) is $\epsilon>0$.
	
	Now consider some bundle $\bundle\in\bundles'$. This bundle is the allocated bundle $\alloci$ for some agent $i$. If this bundle has not been taken by some agent that arrived prior to agent $i$, then since this bundle grants $i$ a utility of $\epsilon$, while other non-empty sets of bundles will grant $i$ a negative utility, agent $i$ is going to take $\alloci$, and get a value of $v_i(\alloci)$ for it. On the other hand, if this bundle was already taken before agent $i$ arrived, by Lemma \ref{lem:demand_structure}, it was taken by some agent $j$ which had a utility of $0$ for this bundle at the end of step \ref{step:merge_bundles}. By the argument above, agent $j$ must have only taken $\alloci$, since every other combination grants $j$ a negative utility. Therefore, the value agent $j$ gets from taking $\alloci$ is $v_j(\alloci)=\price(\alloci)=v_i(\alloci)$. Overall, the obtained welfare is $\sum_i v_i(\alloci)=\OPT$, as desired.
\end{proof}


%% file: k_i_demand_vertex-weighted.tex
\newcommand{\opt}{\text{OPT}}
Let $\agents$ be a set of agents, $\goods$ be a set of items.
A valuation $\val: 2^{\goods} \mapsto \rep$ is \emph{$k$-demand} if 
there exists valuation of the items $w_1: \goods \mapsto \rep$ such that
for any  bundle $B$, $$\val(B)=\max\limits_{X\subseteq B: \abs{X}\leq k}\sum\limits_{b\in X}w_1(b).$$

We say that an item valuation profile $\vals=\{\vali[1],\ldots,\vali[n]\}$ is \textit{$\itemD$} if there exists a function 
$\itemval:\goods\rightarrow\rep$ such that for every agent $i$ and every item $b$, $v_i(b)\in \{0,\itemval(b)\}$.

Finally, we say that a 
valuation profile is \textit{$\ks$-demand $\itemD$} 
for some vector $\ks=(k_1,\ldots,k_n)$ if $\vals$ is $\itemD$ and for every $i$ $v_i$ is $k_i$-demand.\\
Our main result of this section is the following theorem.
\begin{thm}
	\label{thm:k_i-unweighted-demand}
	For any vector $\ks=(k_1,\ldots,k_n)$ and for every valuation profile which is $\ks$-demand $\itemD$ there exists an optimal dynamic bundle-pricing scheme.
\end{thm}

Note that given any optimal allocation $\alloc$, it is possible to construct an optimal allocation $\alloc'$
such that for any agent $i$, the bundle $B_i$ assigned to agent $i$ satisfies $\vali(B_i) = sum_{b \in B_i} w(b)$
by simply removing items of $B_i$ that have non-positive marginal contribution to $\vali(B_i)$.
We call such an allocation a \emph{tight} allocation.

We say that a partition $\bundles_0$ of goods into bundles is a \emph{refinement} of another partition $\bundles_1$ of goods into bundles
if for any two items $u,v$ that belong to a bundle $\bundle_0 \in \bundles_0$, there exists a bundle $\bundle_1 \in \bundles_1$
that contains both $u$ and $v$.

Consider a tight optimal allocation $\alloc$ of bundles to agents.
Let the \emph{relation} graph induced by $\alloc$ be the directed graph $R_{\alloc} = (V,E)$ defined as follows.
For any agent $i$, define $\bundle_i$ be the set of items assigned to agent $i$ and create
vertex $s_i$. 
Create an edge from vertex $s_i$ to vertex $s_j$ in $E$ if $\val_j(\bundle_j) = \val_i(\bundle_j)$.

Let $\epsilon < \min_{u \in \goods}  \val(u)$.
The algorithm at time 0, starts with a bundle for each good.\\For each time $t$, 
Algorithm \pricekdemand~ proceeds as described in Figure~\ref{alg:find-bundle-assignment-vertex-weight}.\\

\begin{figure}[h]
	\MyFrame{		
		\pricekdemand\\
		\textbf{Input:} A set of bundles $\bundles_{t-1}$ and a set of agents $\agents_{t-1}$ and the valuation $v_i: \bundles_{t-1} \rightarrow \rep$ for each agent $i$.\\
		\textbf{Output:} A set of bundles $\bundles_{t}$ such that $\bundles_{t-1}$ is a refinement of $\bundles_t$
                and an assignment of prices $\price_t$ to the bundles of $B_t$.
\begin{enumerate}
\item Compute a tight optimal allocation $\alloc_t$ of the bundles of $\bundles_{t-1}$ to the agents of $N_{t-1}$.
\item For each set $B_i$ of bundles assigned to agent $i$ in $\alloc_t$, create the bundle $B_i$.
\item Construct the relation graph $R_{\alloc_t}$ induced by $\alloc_t$.
\item Remove each edge of $R_{\alloc_t}$ that takes
  part in at least one directed cycle of $R_{\alloc_t}$, this yields a directed acyclic graph $\text{DAG}_t$.
\item Apply a topological sort to $\text{DAG}_t$. It defines an ordering $\sigma$ of the bundles.
\item For any bundle $B_i$ of rank $r$ in $\sigma$, define $\price_t(\bundle_i) \gets \val_i(\bundle_i) - \epsilon^r$.
\item For each bundle $\bundle$ that is not assigned to any agent, define $\price_t(\bundle) \gets \infty$.
\item Return $\{\bundle_0,\ldots,\bundle_n\}$ and $\price_t$
\end{enumerate}

	} \caption{\pricekdemand, a dynamic pricing algorithm for the $\ks$-demand $\itemD$ scenario.}
	
	\label{alg:find-bundle-assignment-vertex-weight}
\end{figure}

We first prove several invariants of the procedure defined in Fig.~\ref{alg:find-bundle-assignment-vertex-weight}.

\begin{lem}
  \label{lem:dag-pricing}
  For any time $t$, for any $s_i,s_j \in \text{DAG}_t$, if there is an edge $\langle s_i,s_j\rangle \in \text{DAG}_t$ then
  $\utili(\bundle_i) > \utili(\bundle_j)$ for the pricing $p_t$.
\end{lem}
\begin{Proof}
  Consider $s_i,s_j  \in \text{DAG}_t$ such that $\langle s_i,s_j\rangle \in \text{DAG}_t$.
  Let $r$ be the rank of $s_i$ in $\sigma$. Since $\langle s_i,s_j\rangle \in \text{DAG}_t$, the rank 
  $r'$ of $s_j$ is higher than $r$, i.e.: $r'>r$.
  
  We write:
  $$\utili(\bundle_i) = \vali(\bundle_i) - p_t(\bundle_i) = \vali(\bundle_i) - \vali(\bundle_i) + \eps^r = \eps^r.$$

  Similarly,
  \begin{equation}
    \label{eq:utiliBj}
    \utili(\bundle_j) = \vali(\bundle_j) - p_t(\bundle_j) = \vali(\bundle_j) - \val_j(\bundle_j) + \eps^{r'}.
  \end{equation}
  Now, recall that there is an edge $\langle s_i,s_t\rangle \in \text{DAG}_t$, and so by definition of $\text{DAG}_t$,
  we have $\vali(\bundle_j) = \val_j(\bundle_j)$. Replacing in Eq.~(\ref{eq:utiliBj}), we have that
  $$\utili(\bundle_j) =  \val_j(\bundle_j) - \val_j(\bundle_j) + \eps^{r'} = \eps^{r'}.$$

  It follows that $\utili(\bundle_i) = \eps^r$ and $\utili(\bundle_j) = \eps^{r'}$.
  Recall $r'>r$ and so $\utili(\bundle_i) > \utili(\bundle_j)$.

\end{Proof}

Define $\out(s_i)$ to be the set $\{s_j \mid \langle s_i, s_j\rangle \in R_{\alloc_t}\}$. Note that
$\out(s_i)$ is defined w.r.t. $R_{\alloc_t}$ and not $\text{DAG}_t$.
We have the following lemma:

\begin{lem}
  \label{lem:ingoing-edge}
  At any time $t$, the arriving agent, say $a_i$, picks bundles in the set $\{\bundle_i\} \cup \out(s_i)$.
\end{lem}
\begin{Proof}
  We first argue that the utility of $\bundle_i$ for agent $i$ is positive. 
  By definition, we have 
  $\utili(\bundle_i) \ge \vali(\bundle_i) - \vali(\bundle_i) + \eps^r = \eps^r$, for 
  some $r \ge 1$. Hence, $\utili(\bundle_i) > 0$.

  We now show that the utility of any $\bundle_j \notin \{\bundle_i\} \cup \out(s_i)$
  for agent $i$ is negative.
  The price of $\bundle_j$ is $\val_j(\bundle_j)- \eps^r$, for some $1\le r \le n$.
  Since there is no edge from $s_i$ to $s_j$ in $R_{\alloc_t}$, we have that 
  $\val_j(\bundle_j) \neq \val_i(\bundle_j)$. Furthermore, observe that the allocation
  is tight. Hence, $\val_j(\bundle_j) = \sum_{b \in \bundle_j} w(b)$ and so, since the 
  valuation are $\itemD$, 
  $\vali(\bundle_j) < \val_j(\bundle_j)$.
  By the choice of $\eps$, it follows that 
  \begin{equation}
    \label{eq:tight}
    \val_i(\bundle_j) + \eps  < \val_j(\bundle_j).
  \end{equation}
  
  By definition of the algorithm, we have 
  $$\utili(\bundle_j) = \vali(\bundle_j) - p_t(\bundle_j) =
  \vali(\bundle_j) - \val_j(\bundle_j) + \eps^r,$$
  for some integer $r \ge 1$.
  Thus, combining with Eq.~(\ref{eq:tight}), 
  we have $\utili(\bundle_j) < 0$. 

  Thus, we have shown that:
  \begin{itemize}
  \item $\utili(\bundle_i) > 0$, and 
  \item $\forall \bundle_j \notin \{\bundle_i\} \cup \out(s_i)$, $\utili(\bundle_j) < 0$.
  \end{itemize}
  We now argue that agent $i$ does not pick
  any bundle $\bundle_j \notin \{\bundle_i\} \cup \out(s_i)$.
  Since $\utili(\bundle_i) > 0$,
  agent $i$ picks a set of bundles that yields positive utility -- in particular the set of
  bundles agent $i$ picks is non-empty. 

  Finally, assume towards contradiction that this set of 
  bundles $B$ contains some $\bundle_j \notin \{\bundle_i\} \cup \out(s_i)$. Consider 
  $B - \{\bundle_j\}$. Since the valuations are $\itemD$ and $\bundle_j$ has negative utility,
  $B - \{\bundle_j\}$ has higher utility than $B$, a contradiction.

  
\end{Proof}

We now proceed to the proof of Theorem \ref{thm:k_i-unweighted-demand}.\\

\begin{Proof}[Proof of Theorem \ref{thm:k_i-unweighted-demand}]
  For any time $t$, let $a_t$ be the agent that arrives at time $t$. 
  Let $\bundles(a_t)$ be the set of bundles that agent $a_{t}$ 
  bought when she arrived at time $t$
  and let $\alloc^{\Alg}_t$ denote the allocation of the bundles defined by $\{(a_1,\bundles(a_1)),\ldots,(a_{t},\bundles(a_{t}))\}$.
  We say that an allocation $\alloc$ extends $\alloc^{\Alg}_t$ if for any $a_j \in \{a_1,\ldots,a_{t}\}$, we have
  that $\alloc$ assigns $\bundles(a_j)$ to $a_j$.

  We aim at proving the following invariant. 
  
  \emph{Invariant: At any time $t$, there exists an optimal allocation $\opt_{t+1}$ which extends $\alloc^{\Alg}_{t}$.}

  Note that the invariant directly implies the theorem
  by taking $t = n$.
  This is true for $t=0$ as no agent has arrived yet.
  We show by induction that it remains true for any time $t>0$.
  Consider a time $t>0$. 
  Let $\alloc_t$ denote the allocation of the bundles of optimal 
  social welfare that (1) extends $\alloc^{\Alg}_{t-1}$ and (2)
  is used by \pricekdemand~ 
  to define the bundles.
  Such an allocation is guaranteed to exist by the induction hypothesis.

  Let $a_t$ be the agent arriving at time $t$.
  For any $i$, let $\bundle_i$ be the bundle assigned to agent $a_i$ in $\alloc_t$.
  Let $\bundles_t$ be the bundles picked by agent $a_t$ when she arrives at time $t$.
  
  We now show that there exists an optimal allocation that extends $\alloc^{\Alg}_{t}$.
  Note that the only difference between $\alloc^{\Alg}_t$ and $\alloc^{\Alg}_{t-1}$
  is that agent $a_t$ gets assigned $\bundles_t$. By the definition of the 
  algorithm, we have $\util_t(\bundle_t)>0$ and so,
  agent $a_t$ picks at least one bundle, i.e.: $\bundles_t \neq \emptyset$.

  We define a new allocation $\alloc^*$ and show that (1) $\alloc^*$ extends $\alloc^{\Alg}_t$ 
  and (2) $\alloc^*$ has optimal social welfare. We start from $\alloc_t$, i.e.: $\alloc^* = \alloc_t$
  and modify $\alloc^*$ in two steps.
  \begin{enumerate}
  \item We make a first modification to $\alloc_t$ if $\bundle_t \notin\bundles_t$.
    If $\bundle_t \notin\bundles_t$, we claim that there exists a
    bundle $\bundle_j$ whose corresponding vertex $s_j$ is such that
    $\langle s_t,s_j\rangle \in R_{\alloc_t}$ and $\langle s_t,s_j\rangle$ was removed at step 3 of
    Procedure \ref{alg:find-bundle-assignment-vertex-weight}.
    By Lemmas~\ref{lem:dag-pricing} and~\ref{lem:ingoing-edge}, we
    have that $\bundles_t \subseteq \{\bundle_t\} \cup \out(s_t)$ and
    if there is an edge $\langle s_t,s_j\rangle$ in the DAG then
    $\util_t(\bundle_j) < \util_t(\bundle_t)$.  Thus, there exists a
    directed cycle $C$ which contains the edge $\langle s_t,s_j\rangle$ in $R_{\alloc_t}$.

    Thus, modify $\alloc^*$ by swapping the bundles along the cycle, namely:
    for each edge
    $\langle s_i,s_{\ell}\rangle$, agent $a_i$ receive bundle
    $\bundle_{\ell}$. Denote by $\bundle^*_j$ the bundle assigned to agent $j$
    in this allocation. Denote by $\alloc^*_1$ the allocation
    obtained after this modification of $\alloc_t$.
    Define $f_{\alloc_t}(\bundle^*_j)$ to be the agent to which $\bundle^*_i$ 
    was assigned to in $\alloc_t$.

  \item Finally, obtain $\alloc^*$ by modifying $\alloc^*_1$ (or $\alloc_t$ if 
    $\bundle_t \in\bundles_t$)  as follows: assign each
    bundle of $\bundles_t$ to agent
    $a_t$ 
    and $\bundle^*_j \setminus \bundles_t$ to any other agent $a_j$.
    For the other agents, the assignment remains the same as in $\alloc^*_1$.
  \end{enumerate}

  By definition, $\alloc^*$ extends $\alloc^{\Alg}_t$. 
  Thus, we only need
  to argue that $\alloc^*$  achieves an optimal social welfare:
  
  $\alloc_t$ achieves optimal social welfare and
  for any edge $\langle s_k,s_{\ell}\rangle \in C$, we have that $\val_k(\bundle_{\ell}) = \val_{\ell}(\bundle_{\ell})$.
  By summing over all $k$ such that $s_k $ is  a vertex of $C$, we obtain that $\SW(\alloc^*_1) = \SW(\alloc_t)$.
  
  We now show that $\SW(\alloc^*) - \SW(\alloc^*_1) \ge 0$.
  Recall that agent $\util_t(\bundles_t) > 0$, so
  for any bundle $\bundle^*_j \in \bundles_t$, $\val_t(\bundle^*_j) - p_t(\bundle^*_j) \ge 0$ as otherwise
  agent $a_t$ would not pick it.
  We thus have:
  \begin{align*}
  \sum_{j \neq t \text{~and~} \bundle^*_j \in \bundles_t} \val_t(\bundle^*_j) &\ge
  \sum_{j \neq t \text{~and~} \bundle^*_j \in \bundles_t} p_t(\bundle^*_j)  
    \\ &\ge 
         \left(\sum_{j \neq t \text{~and~} \bundle^*_j \in \bundles_t}  v_{f_{\alloc_t}(\bundle^*_j)}(\bundle^*_j)
         \right) - n \cdot \eps \\
                                                                              &=
                                                                                \left( \sum_{j \neq t \text{~and~} \bundle^*_j \in \bundles_t} v_j(\bundle^*_j)
  \right) - n \cdot \eps,
  \end{align*}
  and so, by the definition of $\eps$, and since the allocation is tight,
  $$\sum_{j \neq t \text{~and~}\bundle_j \in \bundles_t} \val_t(\bundle^*_j) = 
  \sum_{j \neq t \text{~and~}\bundle_j \in \bundles_t} v_j(\bundle^*_j).$$
  This corresponds to the difference in value for agent $a_t$ in allocation $\alloc^*$.

  Now, for any agent $a_j \neq a_t$ such that $\bundle_j$ is in $\bundles_t$, we have a difference in value in $\alloc^*$ of 
  $- \val_j(\bundle^*_j)$. 
  We combine and sum over all agents and obtain:
  $$\SW(\alloc^*) - \SW(\alloc_t) =
  \sum_{\bundle_j \in \bundles_t \text{~and~}j \neq t}  - \val_j(\bundle^*_j) +
  \sum_{\bundle_j \in \bundles_t \text{~and~}j \neq t}\val_j(\bundle^*_j) = 0.$$

\end{Proof}